\documentclass[onecolumn,aps,prb,longbibliography,showkeys,amsmath,amssymb,floatfix,superscriptaddress,10pt]{revtex4-2}

\usepackage{import}

\usepackage{color,graphics,epsfig,rotating}
\usepackage{graphicx}
\usepackage{dcolumn}
\usepackage{bm}
\usepackage[dvipsnames]{xcolor}

\usepackage{placeins}
\usepackage{mathrsfs}
\usepackage{lmodern}
\usepackage{notoccite}

\usepackage[breaklinks=true]{hyperref}
\usepackage{breakcites}
\usepackage[capitalise]{cleveref}
\usepackage{amsmath}
\usepackage{grffile}
\usepackage{xspace}
\usepackage{xifthen}
\usepackage{soul}
\usepackage{placeins} 
\usepackage{orcidlink}
\usepackage{natbib}
\usepackage{xfrac}
\usepackage{subfigure}

\usepackage[normalem]{ulem}

\begin{document}

{
\onecolumngrid
Notice: This manuscript has been coauthored by UT-Battelle, LLC, under Contract No. DE-AC0500OR22725 with
the U.S. Department of Energy. The United States Government retains and the publisher, by accepting the article for
publication, acknowledges that the United States Government retains a non-exclusive, paid-up, irrevocable, world-wide
license to publish or reproduce the published form of this manuscript, or allow others to do so, for the United States
Government purposes. The Department of Energy will provide public access to these results of federally sponsored
research in accordance with the DOE Public Access Plan (\href{http://energy.gov/downloads/doe-public-access-plan}{http://energy.gov/downloads/doe-public-access-plan}).
}
\title{Directional strain control of magnetism in MnBi$_2$Se$_4$, MnSb$_2$Se$_4$, and MnSb$_2$Te$_4$ freestanding monolayers}
\author{Swarnava Ghosh\,\orcidlink{0000-0003-3800-5264}}
\email{ghoshs@ornl.gov}
\author{Tanvir Sohail \orcidlink{0000-0001-7567-6417}}
\author{Markus Eisenbach\orcidlink{0000-0001-8805-8327}}
\affiliation{National Center for Computational Sciences, Oak Ridge National Laboratory, Oak Ridge, Tennessee 37831, USA}

\date{\today}

\begin{abstract}
MnBi$_2$Se$_4$, MnSb$_2$Se$_4$, and MnSb$_2$Te$4$ are layered magnetic quantum materials of interest for spintronic and topological electronic applications, where control of magnetic order at the atomic scale is essential. These materials are particularly sensitive to strain. We investigate their strain-dependent magnetism in the monolayer limit using first-principles calculations across the full two-dimensional strain space spanned by independent variations of the two in-plane strain components. Our results show that the Mn-projected local moments depend primarily on the volumetric strain, whereas the magnetic ground state, exchange interactions, magnetocrystalline anisotropy, and ordering temperature show pronounced directional and composition-dependent responses. We also observe strain-induced competition between the phases, with MnBi$_2$Se$_4$, and MnSb$_2$Se$_4$ monolayers exhibiting a more diverse range of competing magnetic states, while MnSb$_2$Te$_4$ monolayer remains ferromagnetic over a larger portion of strain space. MnSb$_2$Te$_4$ also exhibits a substantially larger out-of-plane magnetocrystalline anisotropy, while showing a weaker dependence on strain than the Se-based materials. The calculated magnetic ordering temperature maps further highlight the role of two-dimensional strain to tune thermal magnetic stability in these materials. These results establish magnetoelastic trends across this family of Mn-based chalcogenide monolayers.

\end{abstract}
\maketitle


\section{Introduction}

Topological insulators are a class of quantum materials that are insulating in the bulk but have conducting boundary states whose existence is governed by the topology of the electronic band structure \cite{kane2005quantum,fu2007topological,qi2008topological,hasan2010colloquium,qi2011topological}. Introducing magnetism into a topological insulator provides an additional degree of freedom by breaking time-reversal symmetry and coupling the electronic topology with magnetic order. This interplay gives rise to unconventional quantum states and responses, such as the quantum anomalous Hall effect, topological magnetoelectric phenomena, and axion-insulator phases \cite{liu2008quantum,yu2010quantized,mogi2017magnetic,tokura2019magnetic}. Intrinsic magnetic topological materials are particularly attractive because magnetic order and the electronic states responsible for non-trivial topology coexist within the same stoichiometric crystal, even without disorder and inhomogeneity from doping topological insulators with magnetic ions. In these materials, controlling the magnetic ground state, exchange interactions, magnetic anisotropy, and ordering temperature is critical to understanding and utilizing the coupling between magnetism and electronic structure, and to using these materials for next-generation electronic and spintronic applications.

\begin{figure}[h]\centering
\includegraphics[keepaspectratio=true,width=0.9\textwidth]{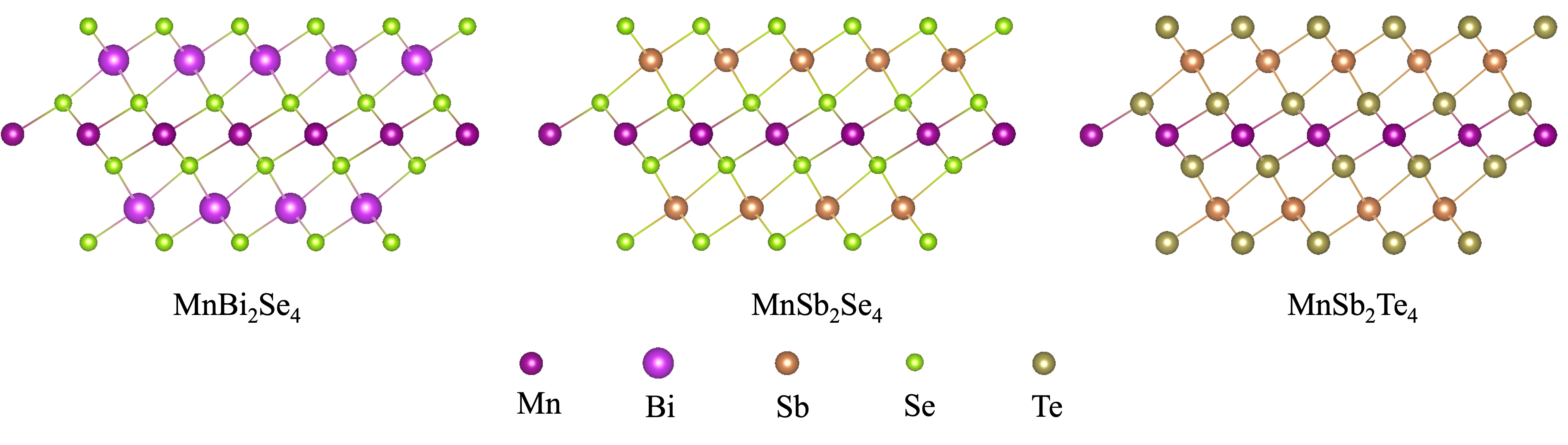}
{\caption{Crystal structures of a single septuple layer of MnBi$_2$Se$_4$,  MnSb$_2$Se$_4$, and MnSb$_2$Te$_4$. Each of these septuple layers is referred to as a monolayer.}\label{Fig:Schematic1}}
\end{figure}

MnBi$_2$Te$_4$ is the first experimentally synthesized intrinsic magnetic topological insulator \cite{li2019intrinsic,gong2019experimental,otrokov2019prediction}. It consists of septuple layers coupled through van der Waals interactions. In the bulk, the Mn moments are ferromagnetically aligned within individual septuple layers and antiferromagnetically coupled between neighboring layers \cite{yan2019crystal,ding2020crystal}. The simultaneous presence of intrinsic magnetism and strong spin--orbit coupling gives rise to a variety of magnetic topological states, including magnetic Dirac surface states, axion-insulator behavior, and the quantum anomalous Hall effect in the few-layer limit \cite{rienks2019large,deng2020quantum,liu2020robust}. These developments have spurred investigation of the ternary telluride family of materials with chemical formula MnA$_2$X$_4$. In this family, MnBi$_2$Se$_4$, MnSb$_2$Se$_4$, and MnSb$_2$Te$_4$ are promising materials whose electronic and magnetic properties have been investigated \cite{chowdhury2019prediction,li2021electronic,zhang2021tunable,zhou2020topological}. Figure \ref{Fig:Schematic1} shows the crystal structures of a single septuple layer, which we refer to as monolayer in our work. These materials show magnetic behavior that differs substantially from the canonical MnBi$_2$Te$_4$ compound. In particular, MnSb$_2$Te$_4$ is sensitive to cation disorder where experiments have shown that Mn--Sb antisite defects can stabilize ferromagnetic or ferrimagnetic interlayer coupling, in contrast to the A-type antiferromagnetism characteristic of MnBi$_2$Te$_4$ \cite{murakami2019realization,wimmer2021mn,riberolles2021evolution}. Epitaxially grown Mn-rich MnSb$_2$Te$_4$ shows out-of-plane ferromagnetism with $T_C\sim45$--$50$ Kelvin and a topological surface state \cite{wimmer2021mn}. MnBi$_2$Se$_4$ has been stabilized experimentally as a van der Waals septuple-layer material using nonequilibrium molecular-beam epitaxy \cite{zhu2021synthesis}, and experiments show A-type antiferromagnetic order with predominantly in-plane moments, distinguishing it from the more Ising-like magnetic character of MnBi$_2$Te$_4$ \cite{zhu2021synthesis,chen2024antiferromagnetic}. First-principles calculations further show that the reduced spin--orbit coupling associated with Se makes MnBi$_2$Se$_4$ more susceptible to pressure- or strain-induced changes in its topological character \cite{li2021electronic,zhang2021tunable}. MnSb$_2$Se$_4$ is distinct from both systems because the experimentally synthesized bulk compound adopts a monoclinic $C2/m$ structure composed of quasi-one-dimensional chains and orders antiferromagnetically near $20$--$22.5$ Kelvin \cite{djieutedjeu2011crystal,kumar2022multiferroicity}. However, first-principles studies have considered a metastable MnBi$_2$Te$_4$-type layered polymorph of MnSb$_2$Se$_4$, for which pressure-tunable magnetic and topological phases have been predicted \cite{zhang2021tunable}. These distinct responses make the three compounds useful for isolating the respective roles of Bi/Sb substitution, Se/Te substitution, spin--orbit coupling, and structure in controlling magnetism within the MnA$_2$X$_4$ family.

Two-dimensional magnetism is of fundamental and technological interest because reduced dimensionality enhances the interplay among magnetic exchange, anisotropy, and thermal fluctuations, enabling strong tunability of collective spin phenomena. As long-range magnetic order is stabilized in atomically thin materials, magnetic states can be tuned through strain, electric fields, interfaces, and chemical substitution. Beyond their fundamental interest, two-dimensional magnets offer opportunities for spintronic, magnetic-memory, and topological functionalities with atomic-scale control. 

In the two-dimensional limit of the MnA$_2$X$_4$ family, the magnetic behavior is governed by intralayer exchange and magnetic anisotropy. This makes monolayers particularly suitable for isolating the effects of chemical composition and strain on magnetic phase stability and for exploring externally tunable magnetism.

Strain provides an effective means of tuning the electronic and optical properties \cite{peng2006strain,ghosh2025strain,guinea2010energy,chang2013orbital,shi2013quasiparticle,peng2020strain}, magnetic response \cite{ahn2026optimizing,christensen2019strain,yang2021strain}, and microstructural phases \cite{ghosh2020influence,ghosh2021precipitation,ghosh2019electronic,sohail2026permutation} without altering chemical composition. In low-dimensional materials, strain can be particularly effective because of their large elastic compliance and strong coupling between lattice, electronic, and magnetic degrees of freedom. Consequently, strain engineering has emerged as a versatile approach for controlling and designing functional properties in quantum and magnetic materials. First-principles studies of two-dimensional magnetetic materials show that strains cause magnetic phase transitions and change the electronic structure through changes in metal--ligand bond lengths, bond angles, and orbital hybridization \cite{webster2018strain,zheng2018tunable,wu2019strain,pizzochero2020inducing,webster2018spinlattice}. Furthermore, strain-dependent changes in the magnetic ordering temperature in suspended two-dimensional magnetic heterostructures have also been reported\cite{siskins2022nanomechanical}. 

The MnA$_2$X$_4$ family is particularly sensitive to strain. Recent first-principles calculations on monolayer MnBi$_2$X$_4$ ($X=$ Te, Se, S) show that biaxial and uniaxial strain influence the Mn magnetic moment differently and can simultaneously alter chemical bonding, magnetic anisotropy, optical response, quantum weight, and mechanical failure behavior \cite{hung2025strain}. Additionally, MnBi$_2$Te$_4$ attached to a ferroelectric substrate leads to polarization-dependent changes in electronic character and magnetic anisotropy, while the combined action of ferroelectric polarization and biaxial strain can enhance the calculated $T_{\mathrm C}$ \cite{xue2020control}. Strain and pressure also modify the electronic topology of MnA$_2$X$_4$ family of compounds \cite{li2021electronic,zhang2021tunable,zhou2020topological}. Collectively, these studies demonstrate the strong coupling between lattice deformation, electronic structure, and magnetism in this family.

Previous studies have established that strain can strongly modify the magnetic and electronic properties of MnBi$_2$Te$_4$ and related compounds, using uniaxial, equibiaxial, and, more recently, two-dimensional strain spaces \cite{li2021electronic,ahn2026optimizing,hung2025strain,xue2020control}. In particular, independent variation of $\varepsilon_{xx}$ and $\varepsilon_{yy}$ can distort crystallographically distinct bonding and exchange pathways differently. This distinction is important for magnetism because strain-induced changes in Mn--Mn and Mn--$X$ bond lengths, Mn--$X$--Mn bond angles, crystal-field splitting, and Mn $3d$--$X$ $p$ hybridization need not evolve identically under deformation along different in-plane directions. Consequently, strain states with similar two-dimensional volumetric strain, $\varepsilon_{xx}+\varepsilon_{yy}$, can exhibit substantially different exchange interactions, magnetic ground states, and ordering temperatures. While the two-dimensional strain dependence of monolayer MnBi$_2$Te$_4$ has recently been examined in detail \cite{ahn2026optimizing}, a systematic understanding of how the full $(\varepsilon_{xx},\varepsilon_{yy})$ strain state controls local moments, competing magnetic configurations, exchange interactions, magnetic anisotropy, and finite-temperature ordering across the chemically related MnBi$_2$Se$_4$, MnSb$_2$Se$_4$, and MnSb$_2$Te$_4$ monolayers remains limited. Such a comparison is needed to distinguish magnetic responses governed primarily by overall lattice dilation or compression from those arising from directional strain and chemical composition.

In this work, we use first-principles calculations to systematically investigate the strain-dependent magnetic properties of freestanding MnBi$_2$Se$_4$, MnSb$_2$Se$_4$, and MnSb$_2$Te$_4$ monolayers over a two-dimensional in-plane strain space. We first study the influence of the Hubbard $U$ parameter in DFT+$U$ calculations of these materials. Next, the magnetic ground-state phase diagrams are determined directly by comparing the DFT+$U$ total energies of the ferromagnetic and three antiferromagnetic configurations at each strain state. These energies are subsequently mapped onto a Heisenberg model to provide a microscopic description of the evolution of the competing exchange interactions. We also determine the strain dependence of the Mn-projected local magnetic moments, magnetic exchange parameters, and magnetocrystalline anisotropy and use Monte Carlo simulations to evaluate the corresponding magnetic ordering temperatures.

\section{Methods}
We performed first principles density functional theory calculations \cite{hohenberg1964inhomogeneous,kohn1965self} using the Vienna {\it{ ab initio}} simulation package (VASP) \cite{kresse1993ab,kresse1996efficient} for computing the ground state energies and magnetic properties of  MnBi$_2$Se$_4$,  MnSb$_2$Se$_4$,  and MnSb$_2$Te$_4$. We used a plane-wave cutoff of $400$ eV with projector augmented wave method (PAW) pseudopotentials \cite{blochl2003projector} and employed an $15\times15\times1$ $\Gamma$-centered $k$-points sampling. Electronic self-consistency was achieved using an energy convergence threshold of $10^{-8}$ eV. As exchange-correlation functional, generalized gradient approximation (GGA) of Perdew-Burke-Ernzerhof (PBE) \cite{perdew1998perdew} was used. We set the Hellman-Feynman force criterion to $0.01$ eV/Angstrom for structural and atomic relaxation. Mn has partially filled $d$-orbitals, and their electronic behavior can be intricate, involving strong electron-electron interactions. Hence, in such cases, the introduction of Hubbard $U$ in the framework of DFT is necessary to account for the on-site Coulomb repulsion between electrons in the transition metal $d$-orbitals. We have used $U$ parameter values between $0$ eV and $6$ eV on Mn 3$d$ orbitals. 

To investigate the effects of mechanical deformation, in-plane tensile and compressive strains were applied by systematically scaling the equilibrium lattice parameters. During the structural optimization of each strained configuration, the lattice vectors were kept fixed while the internal atomic positions were fully relaxed until the residual forces on each atom were below $0.01$ eV/Angstrom.

To calculate the magnetic ordering temperature $T_C$, we first fit a classical Heisenberg model using exchange constants and magnetic anisotropy constants calculated using DFT. We then use this Hamiltonian to perform Metropolis Monte Carlo simulations with single-spin moves using the OWL Monte-Carlo package \cite{Li2021} on a $16\times 16$ two-dimensional hexagonal lattice. For each simulation, we performed 100,000 thermalization steps and $1,600,000$ measurement steps with $256$ spin moves for each step.
 
 \section{Results and discussion}

\subsection{Effect of Hubbard $U$ parameter}\label{Sec:U}
Previous DFT+$U$ calculations of this family of compounds have employed $U$ values ranging from $2$ to $5$ eV \cite{liu2020robust,zhang2021tunable,zhou2020topological,riberolles2021evolution,liu2021site,alfonsov2021strongly}. However, the calculated electronic and magnetic properties of these materials are highly sensitive to the treatment of Mn $3d$ correlations, particularly to the choice of the Hubbard $U$ parameter. We therefore first examine the dependence of the relative magnetic-state energies, Mn local moments, and exchange parameters on $U$ before selecting the value used in the subsequent strained calculations.

\begin{figure}[h]\centering
\includegraphics[keepaspectratio=true,width=0.35\textwidth]{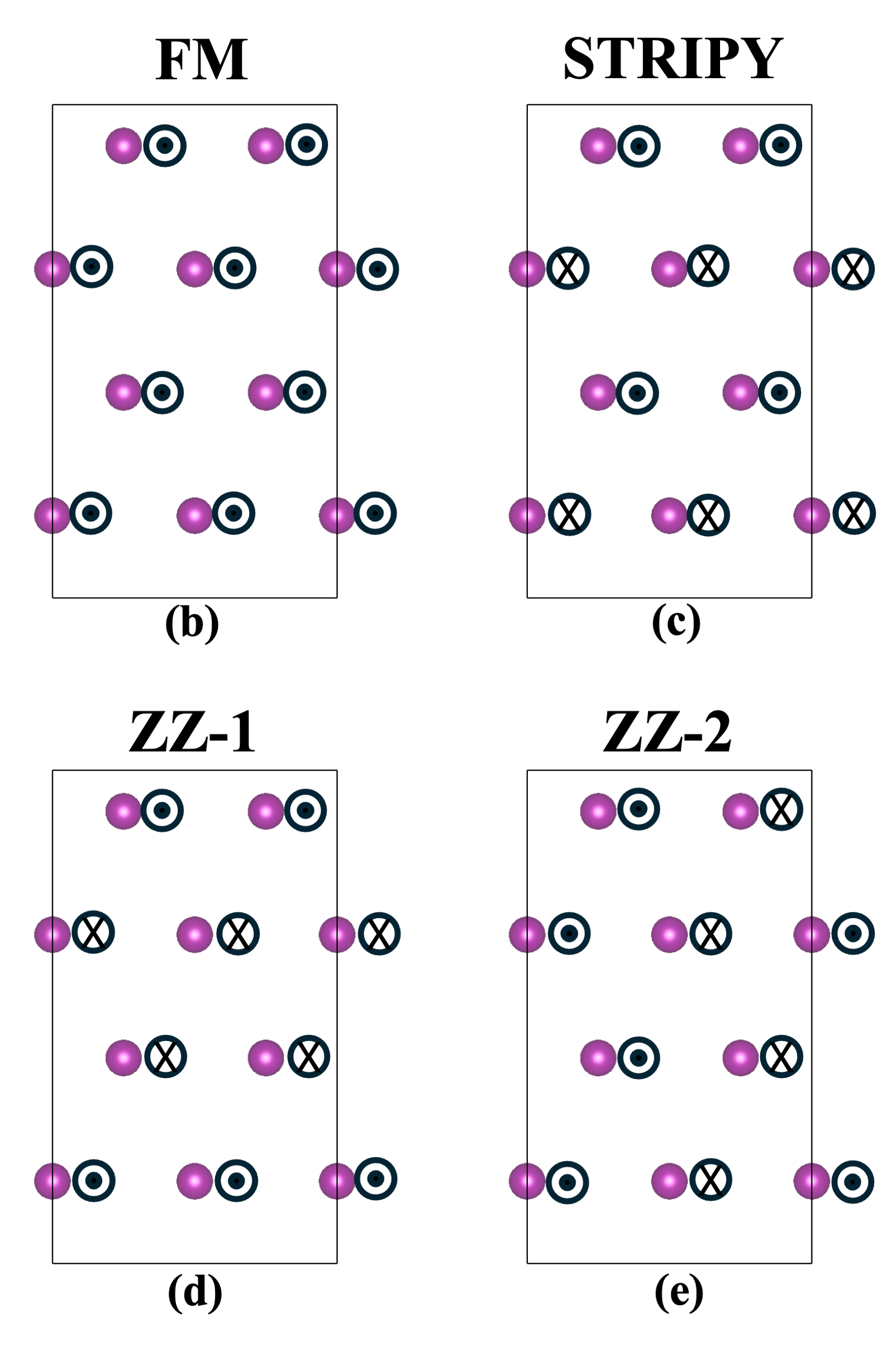}
{\caption{Schematic representation of the four magnetic configurations considered in this work. Only the plane containing the Mn atoms is shown. Circles containing black dots and crosses denote magnetic moments pointing out of and into the plane of the page, respectively.}\label{Fig:MagneticConfig}}
\end{figure}

Four different magnetic configurations as shown in Figure  \ref{Fig:MagneticConfig} are considered. They are ferromagnetic (FM) and three different antiferromagnetic (AFM) configurations forming a stripy pattern, and two zigzag orderings of spin-up and spin-down Mn moments.

\begin{figure}[h]\centering
\subfigure[MnBi$_2$Se$_4$]{\includegraphics[keepaspectratio=true,width=0.31\textwidth]{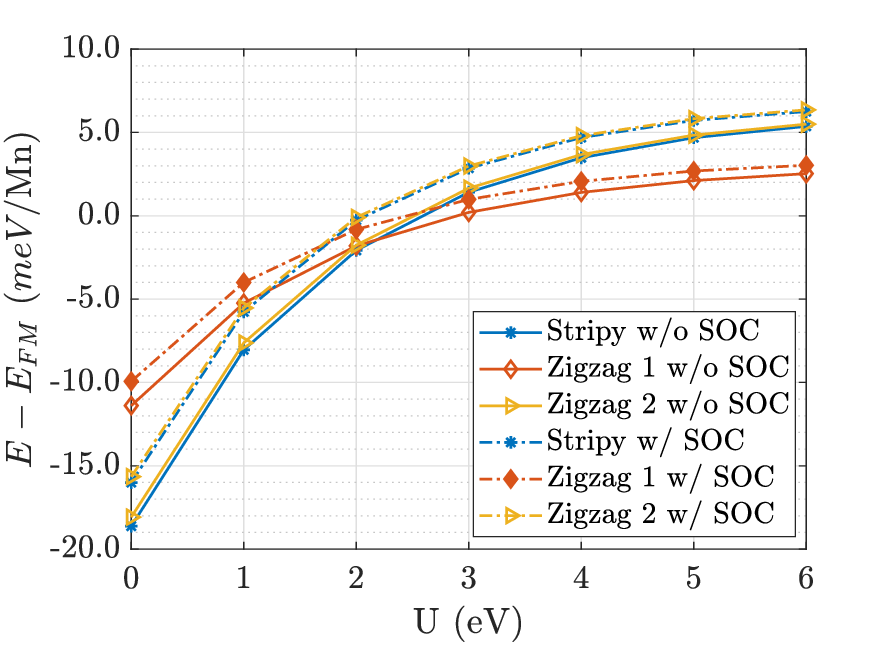}\label{Fig:MBS:energy}}
 \hspace{-1em}
\subfigure[MnSb$_2$Se$_4$]{\includegraphics[keepaspectratio=true,width=0.3\textwidth]{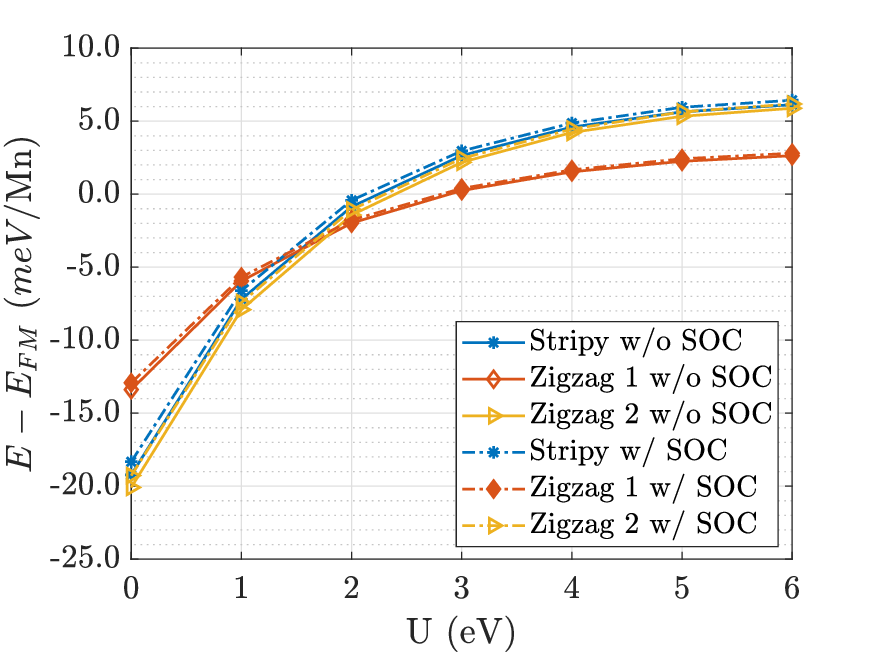}\label{Fig:MSS:energy}}
 \hspace{-1em}
\subfigure[MnSb$_2$Te$_4$]{\includegraphics[keepaspectratio=true,width=0.3\textwidth]{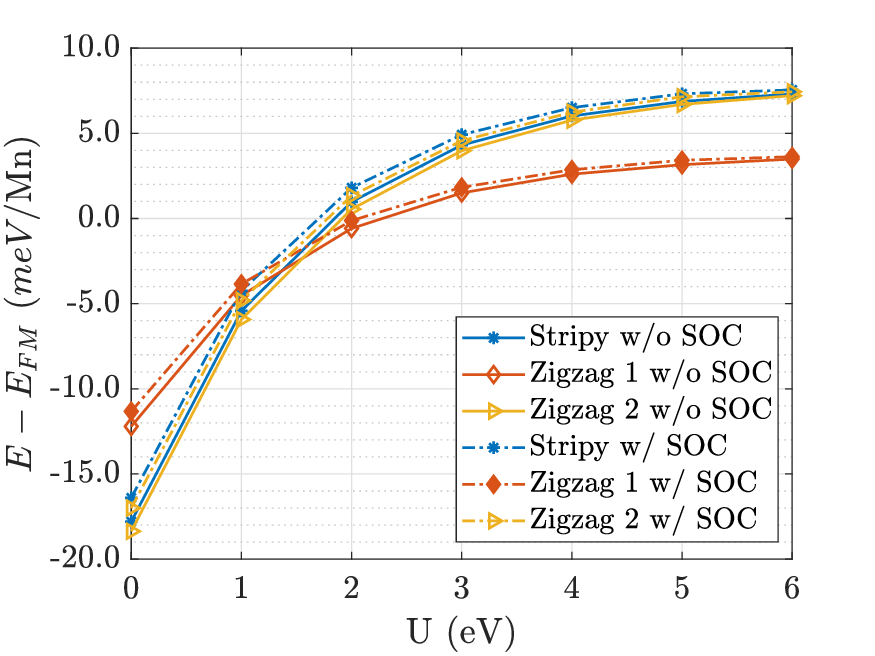}\label{Fig:MST:energy}}
{\caption{The difference in energy between the three antiferromagnetic states and the ferromagnetic states versus the Hubbard $U$ parameter. Calculations without SOC are shown as solid lines and calculations with SOC are shown as dash-dotted lines, respectively. }\label{Fig:EnergyvsU}}
\end{figure}

Figure \ref{Fig:EnergyvsU} shows the change in the energy difference between the ferromagnetic (FM) state and the three antiferromagnetic (AFM) configurations -- Stripy, ZZ-1, and ZZ-2 -- as shown in Figure \ref{Fig:MagneticConfig}, as a function of the Hubbard $U$ parameter. Positive energy differences indicate that the FM state is energetically favored over the corresponding AFM configuration. Calculations were performed both with and without spin–orbit coupling (SOC).

At $U=0$ eV, where no Hubbard correction is applied, the AFM configurations are lower in energy than the FM state. The AFM states remain energetically favorable for small $U$ values up to approximately $2$ eV. With increasing $U$, however, the relative stability changes, and for $U\gtrsim3$ eV the FM state becomes the lowest-energy configuration among the magnetic states considered. The inclusion of SOC produces only minor changes in the relative energies and does not alter the overall $U$-dependent trend. Furthermore, the energy differences vary only weakly between $U=5$ and $6$ eV, suggesting that the relative magnetic energetics become less sensitive to further increases in $U$ in this regime.

\begin{figure}[h]\centering
\subfigure[MnBi$_2$Se$_4$]{\includegraphics[keepaspectratio=true,width=0.3\textwidth]{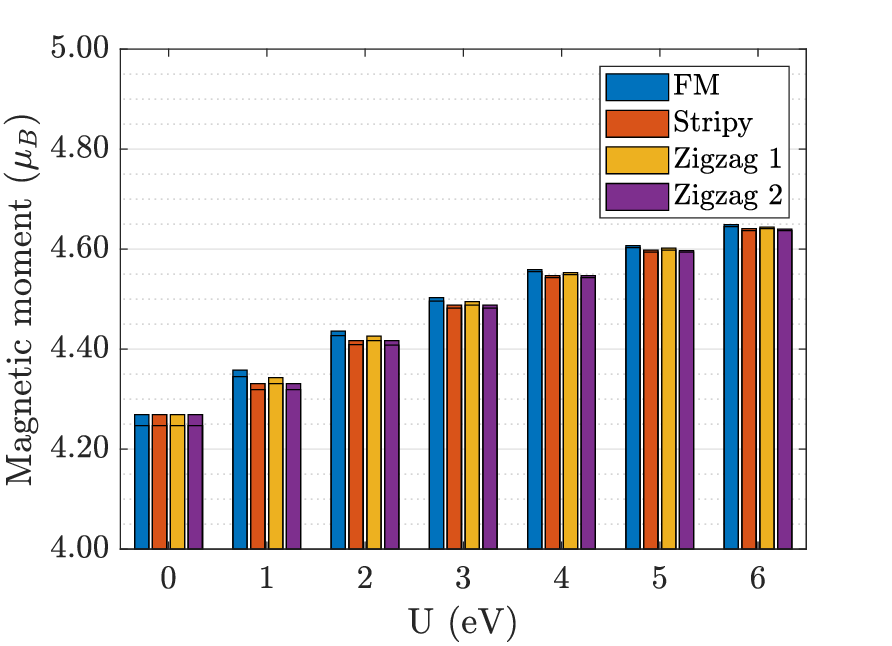}\label{Fig:MBS:moment}}
 \hspace{-1em}
\subfigure[MnSb$_2$Se$_4$]{\includegraphics[keepaspectratio=true,width=0.3\textwidth]{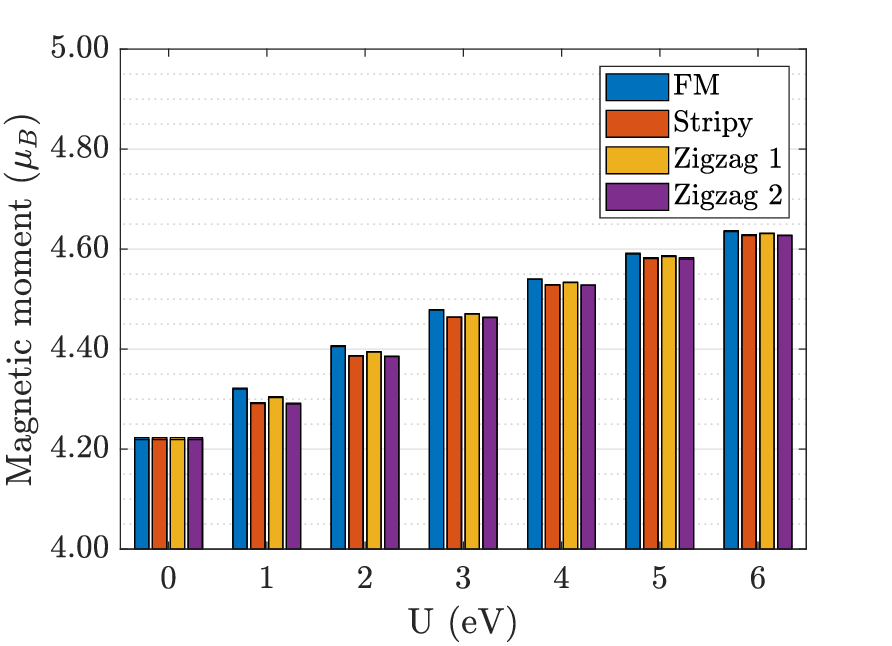}\label{Fig:MSS:moment}}
 \hspace{-1em}
\subfigure[MnSb$_2$Te$_4$]{\includegraphics[keepaspectratio=true,width=0.3\textwidth]{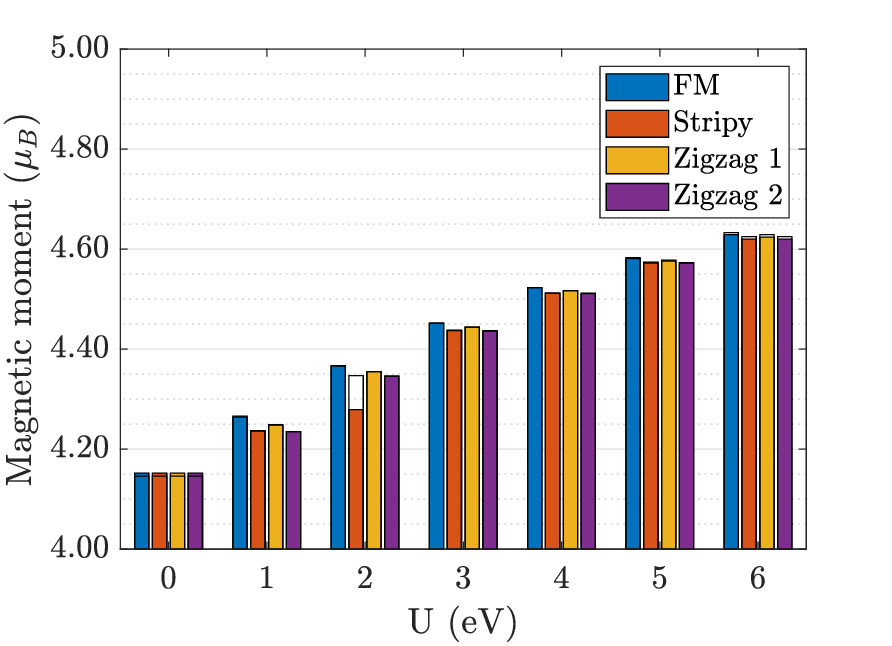}\label{Fig:MST:moment}}
{\caption{Effect of Hubbard $U$ parameter on the magnetic moments. Calculations with SOC are shown as filled bars whereas calculations without SOC are shown as unfilled bars.}\label{Fig:MomentvsU}}
\end{figure}

Figure \ref{Fig:MomentvsU} shows the local magnetic moment on the Mn atoms as a function of the Hubbard $U$ parameter. For a given value of $U$, the calculated local magnetic moments are nearly identical for the FM, Stripy, ZZ-1, and ZZ-2 configurations. This weak dependence on magnetic ordering indicates that the magnitude of the Mn moment is governed predominantly by the local electronic structure and intra-atomic exchange interactions, whereas the different magnetic configurations only change the relative orientation of the Mn moments through intersite exchange interactions. Therefore, although the four magnetic states can differ appreciably in total energy, as shown in Figure \ref{Fig:EnergyvsU}, the magnitude of the local moment on each Mn site remains largely unchanged. This distinction highlights that the energetic competition between the FM and AFM configurations is controlled by the interactions between neighboring local moments rather than by changes in the magnitude of the moments themselves.

The inclusion of spin–orbit coupling (SOC) produces only a negligible change in the local magnetic moment on the Mn atoms. This is consistent with the comparatively weak SOC of Mn $3d$ states and the predominantly spin-derived character of the Mn magnetic moment. In addition, the orbital contribution to the magnetic moment is quenched by the local crystal-field environment. Spin-orbit coupling therefore affects the orientation of the magnetic moments and the associated magnetic anisotropy more strongly than their magnitude.

In contrast, the local magnetic moment on the Mn atoms increases systematically with increasing $U$. The Hubbard correction enhances the localization of the Mn $3d$ electrons by reducing their tendency to delocalize through hybridization with neighboring atoms. Increased localization strengthens the spin polarization of the Mn $3d$ states and increases the occupation imbalance between the majority- and minority-spin channels, resulting in a larger local magnetic moment. The increase in the Mn moment with $U$ therefore reflects the progressive localization and enhanced spin polarization of the Mn $3d$ electrons.

The exchange parameters $J_i$ between Mn atoms were determined by mapping the DFT total energies of the four magnetic configurations shown in Figure \ref{Fig:MagneticConfig} onto a classical Heisenberg spin Hamiltonian,

\begin{equation}\label{Eq:HeisenbergHamiltonian}
    H=-J_1 \sum_{(i,j)\in \mathcal{N}_1} \bm{S}_i \cdot \bm{S}_j  -J_2 \sum_{(i,j)\in \mathcal{N}_2} \bm{S}_i\cdot \bm{S}_j -J_3 \sum_{(i,j)\in \mathcal{N}_3} \bm{S}_i\cdot \bm{S}_j \,
\end{equation}
where $\bm{S}_i$ denotes the spin on site $i$, and $J_1$, $J_2$, and $J_3$ are the first-, second-, and third-nearest-neighbor exchange parameters, respectively. The sets $\mathcal{N}_1$, $\mathcal{N}_2$, and $\mathcal{N}_3$ contain the corresponding first-, second-, and third-nearest-neighbor Mn--Mn pairs. With the sign convention adopted in Equation \ref{Eq:HeisenbergHamiltonian}, positive values of $J_i$ favor ferromagnetic alignment, whereas negative values favor antiferromagnetic alignment. For a two-dimensional magnetic system, magnetic anisotropy is required to stabilize finite-temperature long-range order, and is included explicitly in the Monte Carlo simulations used to determine the ordering temperature (see Equation \ref{Eq:HamiltonianWithAnisotropy}). By contrast, the exchange parameters $J_i$ are extracted from the relative DFT energies of the collinear magnetic configurations and are obtained from the exchange part of the spin Hamiltonian alone.

For the expanded unit cell used to construct the four magnetic configurations, the corresponding Heisenberg energies can be written as \cite{xue2020control}

\begin{eqnarray}
\label{Eq:FourEqns}
E^{\mathrm{FM}} &=& E_0 -24J_1|S|^2 -24J_2|S|^2 -24J_3|S|^2, \nonumber\\
E^{\mathrm{Stripy}} &=& E_0 +8J_1|S|^2 +8J_2|S|^2 -24J_3|S|^2, \nonumber\\
E^{\mathrm{ZZ1}} &=& E_0 -8J_1|S|^2 +8J_2|S|^2 +8J_3|S|^2, \nonumber\\
E^{\mathrm{ZZ2}} &=& E_0 +8J_1|S|^2 -8J_2|S|^2 +8J_3|S|^2 .
\end{eqnarray}

Here, $E_0$ represents the spin-independent contribution to the total energy. By equating these expressions to the corresponding DFT total energies, the resulting system of four equations was solved simultaneously to obtain $J_1$, $J_2$, $J_3$, and $E_0$.

\begin{figure}[h]\centering
\subfigure[MnBi$_2$Se$_4$]{\includegraphics[keepaspectratio=true,width=0.3\textwidth]{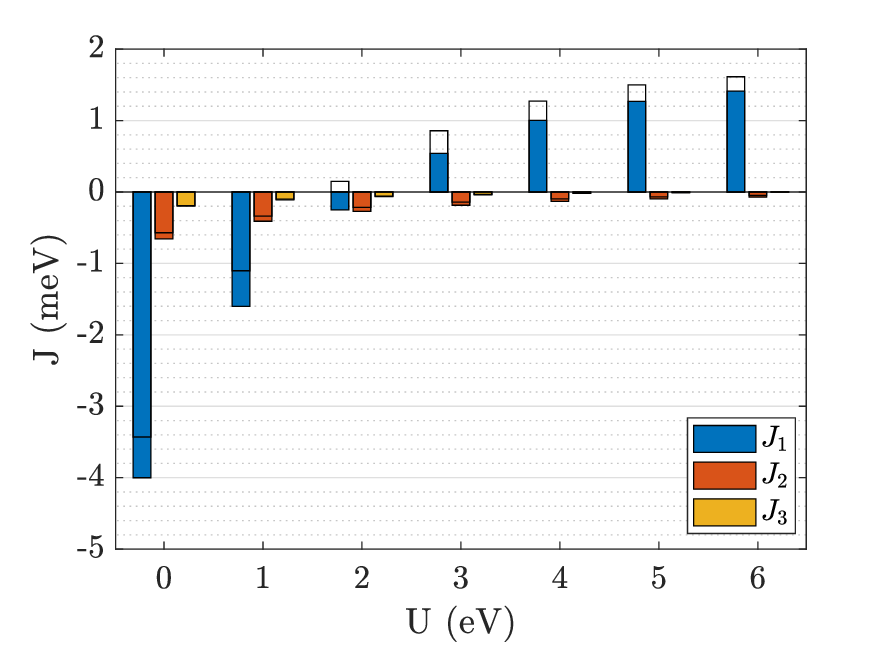}\label{Fig:MBS:J}}
 \hspace{-1em}
\subfigure[MnSb$_2$Se$_4$]{\includegraphics[keepaspectratio=true,width=0.3\textwidth]{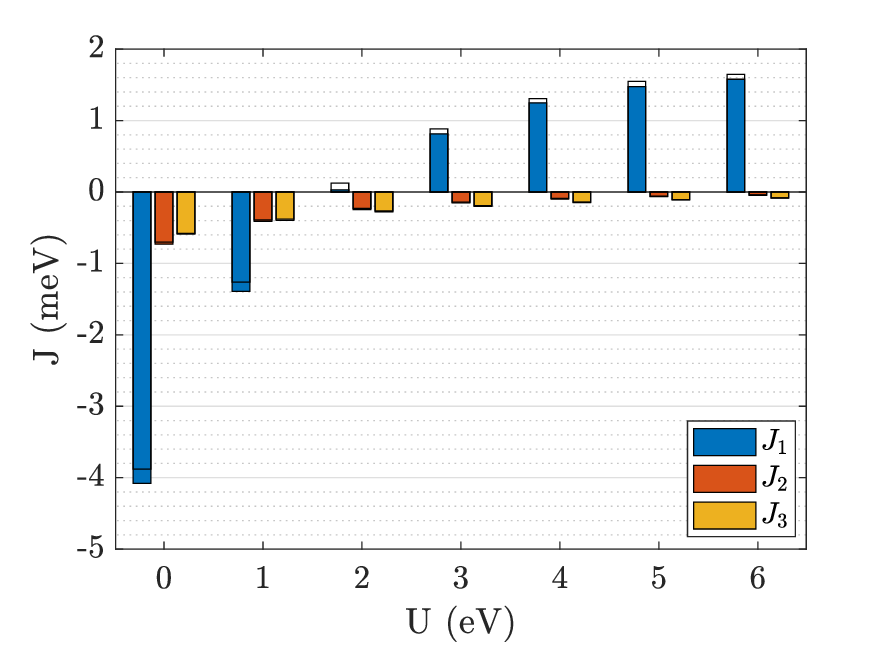}\label{Fig:MSS:J}}
 \hspace{-1em}
\subfigure[MnSb$_2$Te$_4$]{\includegraphics[keepaspectratio=true,width=0.3\textwidth]{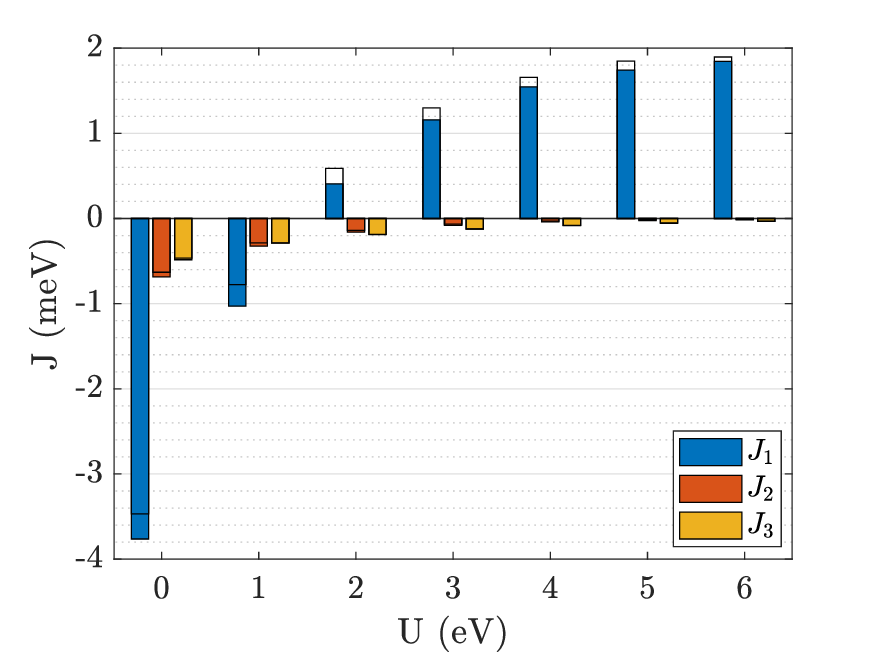}\label{Fig:MST:J}}
{\caption{Effect of Hubbard $U$ parameter on magnetic exchange constants $J_1$, $J_2$ and $J_3$. Calculations with SOC are shown as filled bars, whereas calculations without SOC are shown as unfilled bars.}\label{Fig:JvsU}}
\end{figure}

Figure \ref{Fig:JvsU} shows the calculated first-, second-, and third-nearest-neighbor exchange parameters, $J_1$, $J_2$, and $J_3$, respectively, as a function of the Hubbard $U$ parameter over the range $0$--$6$ eV. The magnitude of $J_1$ is substantially larger than those of $J_2$ and $J_3$ throughout the range considered, indicating that the nearest-neighbor Mn--Mn interaction provides the dominant contribution to the magnetic energetics and therefore plays the primary role in determining the magnetic ground state. The comparatively weaker $J_2$ and $J_3$ interactions are consistent with the reduced strength of exchange interactions between more distant Mn sites, although their contributions may still influence the delicate balance between competing magnetic configurations.

At $U=0$ eV, corresponding to calculations without a Hubbard correction, $J_1$, $J_2$, and $J_3$ are all negative. Within the sign convention adopted in Equation \ref{Eq:HeisenbergHamiltonian}, these negative values correspond to antiferromagnetic exchange interactions and are consistent with the lower energies obtained for the antiferromagnetic configurations relative to the ferromagnetic state. Upon increasing $U$, however, a pronounced change is observed in $J_1$, namely, the nearest-neighbor interaction switches from antiferromagnetic at $U=0$ eV to ferromagnetic for $U\geq3$ eV. This crossover is consistent with the corresponding change in the calculated magnetic ground state discussed above, where the FM configuration becomes energetically favorable at larger $U$.

The strong $U$ dependence of $J_1$ can be understood in terms of the effect of the Hubbard correction on the Mn $3d$ electronic states. Increasing $U$ enhances the localization and spin polarization of the Mn $3d$ electrons and modifies their hybridization with the surrounding ligand states. These changes alter the balance between competing exchange mechanisms that favor antiferromagnetic and ferromagnetic alignment, leading to the observed sign reversal of $J_1$. Because $J_1$ is considerably larger in magnitude than $J_2$ and $J_3$, this crossover in the nearest-neighbor interaction explains the transition from an antiferromagnetic to a ferromagnetic ground state with increasing $U$.

The inclusion of spin--orbit coupling produces only minor changes in the calculated exchange parameters and does not alter these overall trends. This behavior is consistent with the comparatively weak spin--orbit interaction associated with Mn $3d$ states, such that SOC has a limited effect on the dominant isotropic exchange interactions. Its influence is therefore expected to be more pronounced for anisotropic magnetic properties, such as the magnetocrystalline anisotropy, than for the exchange parameters considered here.

\subsection{Effect of strain}\label{Sec:strain}
In this section, we discuss the effect of in-plane strain on the magnetic ground state, magnetic exchange, anisotropy, and magnetic ordering temperature. As studied in Section \ref{Sec:U}, increasing $U$ beyond $5$ eV does not significantly influence the results. So, we choose $U$= $5$ eV for these simulations. In this work, we adopt the convention that negative and positive strain values denote compression and tension, respectively.

\begin{figure}[h]
\centering
\includegraphics[keepaspectratio=true,width=0.8\textwidth]{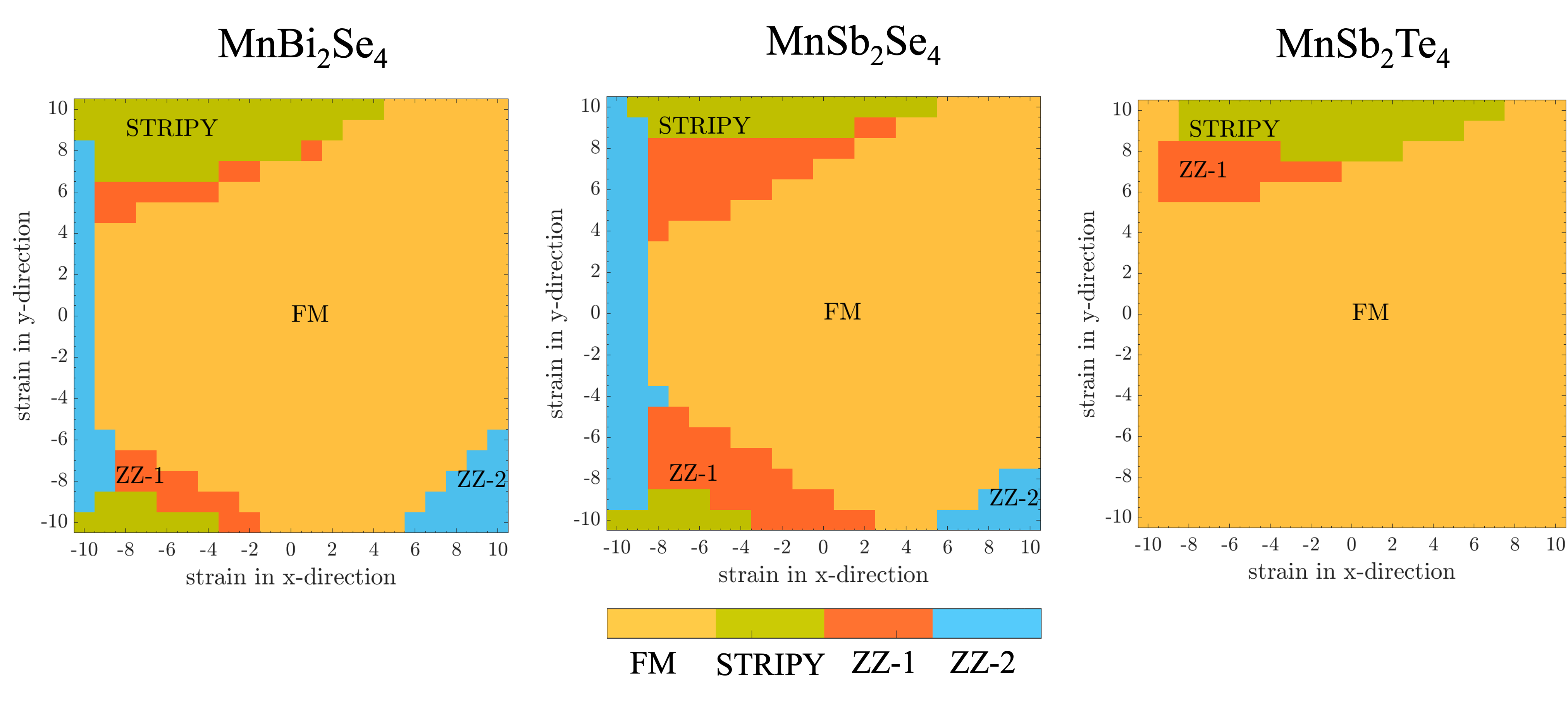}
{\caption{Phase diagram of magnetic ground state with strain ($\%$)}\label{PhaseDiagram}}
\end{figure}

\paragraph{Magnetic phase diagram - }
Figure \ref{PhaseDiagram} summarizes the magnetic ground states of MnBi$_2$Se$_4$, MnSb$_2$Se$_4$, and MnSb$_2$Te$_4$ over the investigated two-dimensional in-plane strain space. We obtained the phase diagrams by comparing the energies of the FM, Stripy, ZZ-1, and ZZ-2 magnetic configurations and labeling the minimum-energy state as the ground-state phase at each combination of $\varepsilon_{xx}$ and $\varepsilon_{yy}$. A common feature of all three monolayers is the large region surrounding the unstrained configuration in which the FM state remains energetically preferred. However, the extent of this FM region and the magnetic phases that emerge at larger strains depend strongly on the chemical composition, demonstrating that the magnetic response to deformation is both anisotropic and material-specific.

For MnBi$_2$Se$_4$, the FM phase occupies most of the central region of the phase diagram and remains stable over a comparatively broad range of moderate strains. Antiferromagnetic phases are seen mainly toward the boundaries of the strain space. At large tensile $\varepsilon_{yy}$, the Stripy phase becomes favorable over a considerable range of compressive $\varepsilon_{xx}$, with the phase boundary shifting progressively towards larger $\varepsilon_{yy}$ as compressive $\varepsilon_{xx}$ is decreased. The ZZ-1 phase occurs predominantly under strongly compressive $\varepsilon_{xx}$ combined with either sufficiently large tensile or compressive $\varepsilon_{yy}$, while the ZZ-2 state appears mainly near the most strongly compressed-$x$ region and in the lower-right portion of the phase diagram at large tensile $\varepsilon_{xx}$ and compressive $\varepsilon_{yy}$. The occurrence of several different antiferromagnetic configurations at the boundaries of the diagram indicates that large anisotropic distortions modify the strengths of exchange interactions sufficiently to destabilize the FM state.

MnSb$_2$Se$_4$ shows a qualitatively similar, but more extensive, competition between the magnetic phases. The FM state again dominates the central strain region, whereas the Stripy, ZZ-1, and ZZ-2 configurations become favorable at larger deformations. In particular, the Stripy phase is stabilized primarily at large tensile $\varepsilon_{yy}$, while a broad ZZ-1 region develops for compressive $\varepsilon_{xx}$ over both positive and negative $\varepsilon_{yy}$. The phase boundaries are generally shifted relative to those of MnBi$_2$Se$_4$, indicating that replacing Bi with Sb gives rise to several closely competing AFM phases under large strains.

A markedly different behavior is observed for MnSb$_2$Te$_4$. The FM phase occupies almost the entire strain range investigated, demonstrating considerably greater stability of ferromagnetic ordering against in-plane deformation than in the two Se-based monolayers. Antiferromagnetic ordering is confined mainly to the region of large tensile $\varepsilon_{yy}$. Within this region, the ZZ-1 state is favored primarily when tensile $y$-strain is combined with compressive $\varepsilon_{xx}$, whereas the Stripy state becomes stable at even larger tensile $\varepsilon_{yy}$ and extends over a wider range of $\varepsilon_{xx}$. Notably, the ZZ-2 phase is not the ground state anywhere within the strain window considered for MnSb$_2$Te$_4$.

The strong directional character of the phase boundaries further demonstrates that the magnetic ground state is controlled by the full strain tensor rather than by the scalar strain magnitude or the two-dimensional volumetric strain alone. Deformation along the $x$ and $y$ directions modifies different Mn--$X$ bond lengths, Mn--$X$--Mn bond angles, and Mn $3d$--$X$ $p$ hybridization pathways ($X=$ Se or Te), causing the individual exchange interactions to respond differently to $\varepsilon_{xx}$ and $\varepsilon_{yy}$.

The comparison among the three materials also reveals a clear chemical trend. The two Se-based monolayers exhibit richer strain-induced magnetic phase behavior, with all four magnetic configurations appearing within the investigated strain window. By contrast, MnSb$_2$Te$_4$ maintains ferromagnetic ordering over most of the same strain range and exhibits only localized Stripy and ZZ-1 regions at large tensile $\varepsilon_{yy}$. This difference indicates that chemical substitution modifies not only the absolute exchange strengths but also their sensitivity to lattice deformation. In particular, the more spatially extended Te $p$ states and the resulting Mn $3d$--Te $p$ hybridization can produce an exchange network that responds differently to structural distortions than the corresponding Se-mediated interactions.

Overall, Figure \ref{PhaseDiagram} demonstrates that strain provides an effective route for controlling the magnetic ground state of these monolayers and emphasizes that magnetic-phase engineering in this family requires control over both the magnitude and direction of the applied strain. Interestingly, the strain-dependent magnetic phase diagram obtained here for MnSb$_2$Te$_4$ using a fixed Hubbard parameter exhibits qualitative similarities to the Diffusion Monte Carlo (DMC)-informed phase diagram reported for monolayer MnBi$_2$Te$_4$, in which $U$ was optimized as a function of strain \cite{ahn2026optimizing}. This agreement suggests that some of the dominant strain-induced changes in magnetic phase stability are similar across the Te-based members of the Mn$A_2X_4$ family.

The present calculations describe freestanding monolayers and therefore establish the intrinsic magnetoelastic response of MnBi$_2$Se$_4$, MnSb$_2$Se$_4$, and MnSb$_2$Te$_4$ in the absence of substrate-induced interactions. These results provide a basis for exploring strain-controlled device architectures, for example by transferring strain from piezoelectric, ferroelectric, or flexible substrates. Furthermore, when the monolayers are supported on a substrate, interfacial hybridization, charge transfer, electrostatic fields, and substrate-induced symmetry breaking can modify the exchange interactions and magnetic anisotropy in addition to mechanical strain.

\begin{figure}[h]\centering
\subfigure[MnBi$_2$Se$_4$]{\includegraphics[keepaspectratio=true,width=0.3\textwidth]{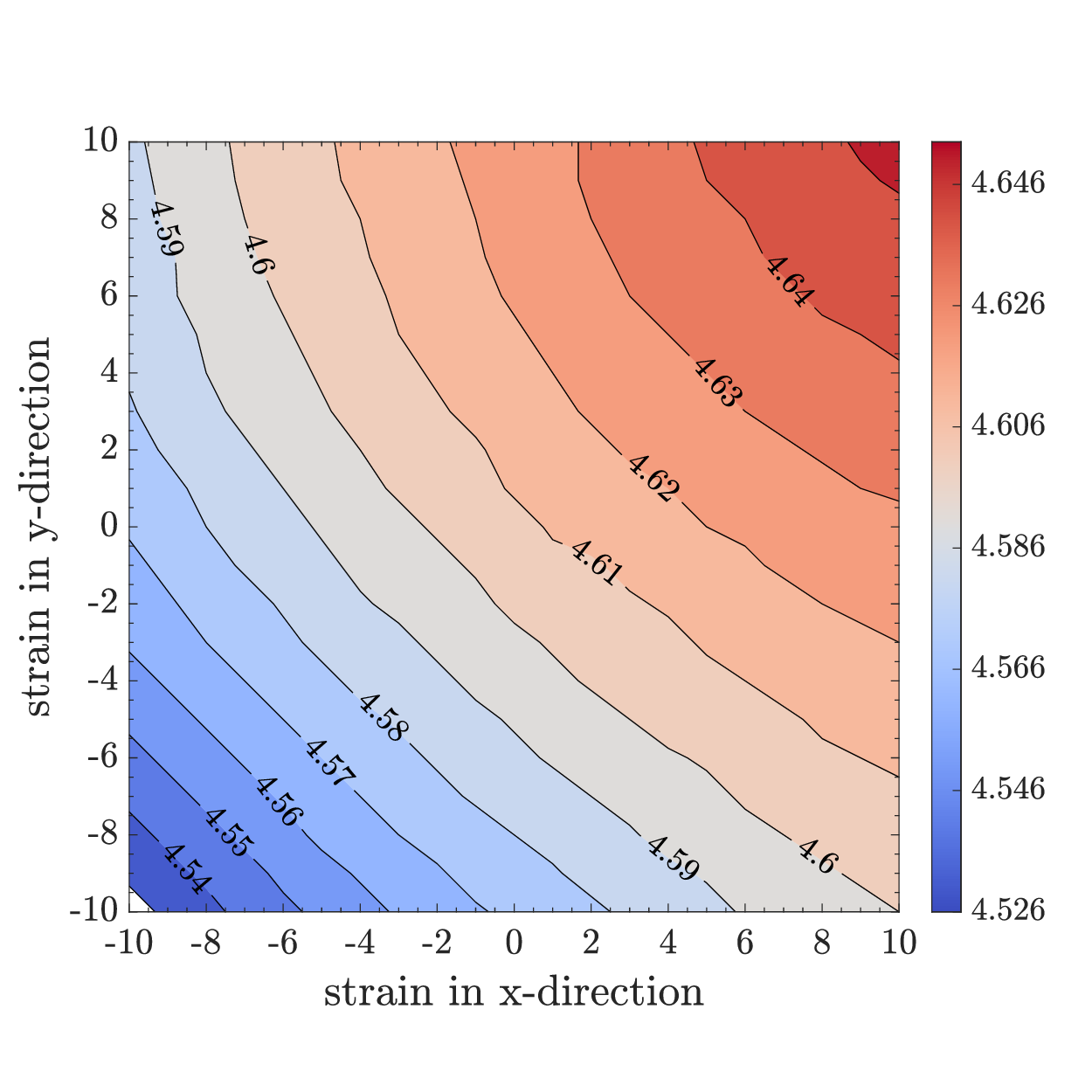}}
 \hspace{-1em}
\subfigure[MnSb$_2$Se$_4$]{\includegraphics[keepaspectratio=true,width=0.3\textwidth]{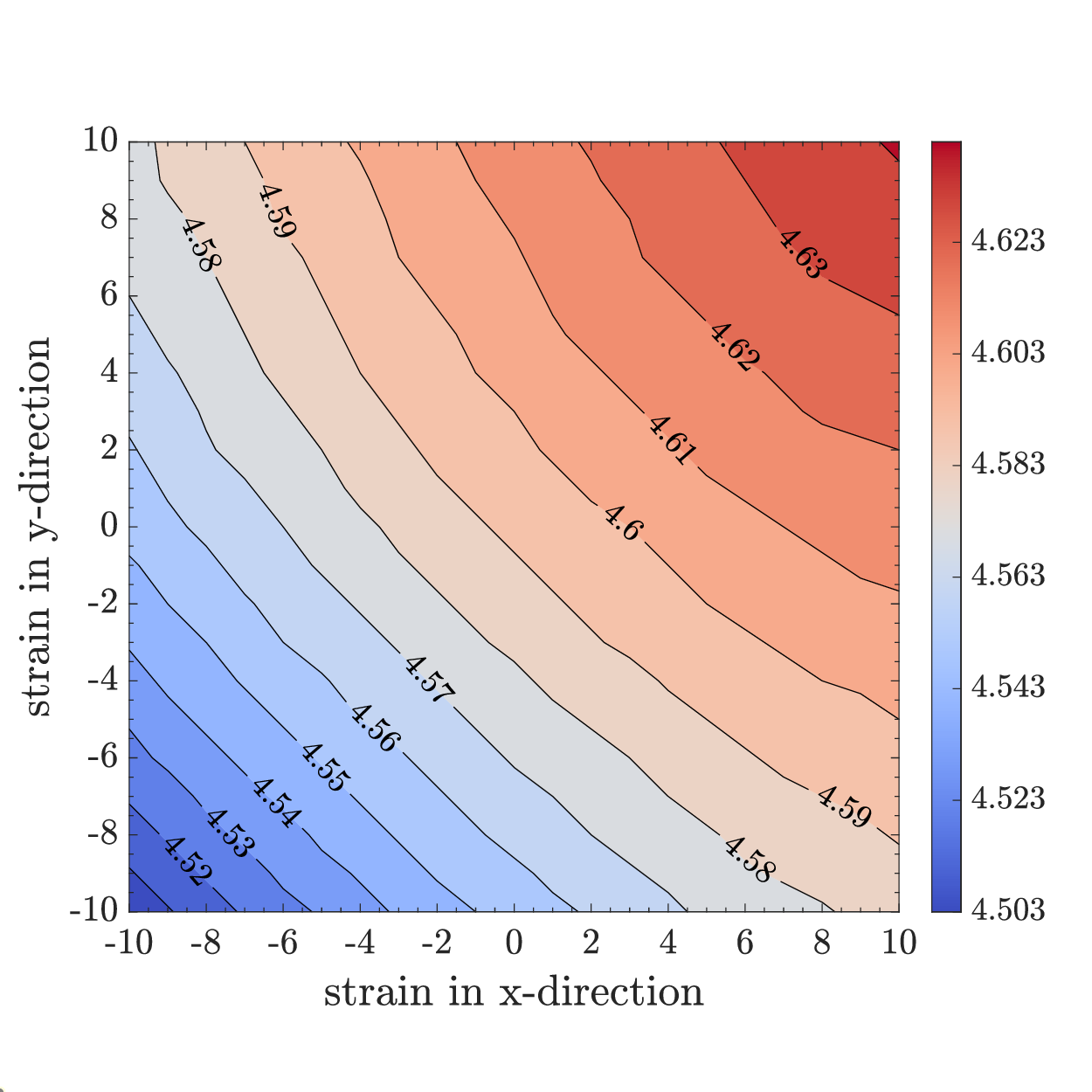}}
 \hspace{-1em}
\subfigure[MnSb$_2$Te$_4$]{\includegraphics[keepaspectratio=true,width=0.3\textwidth]{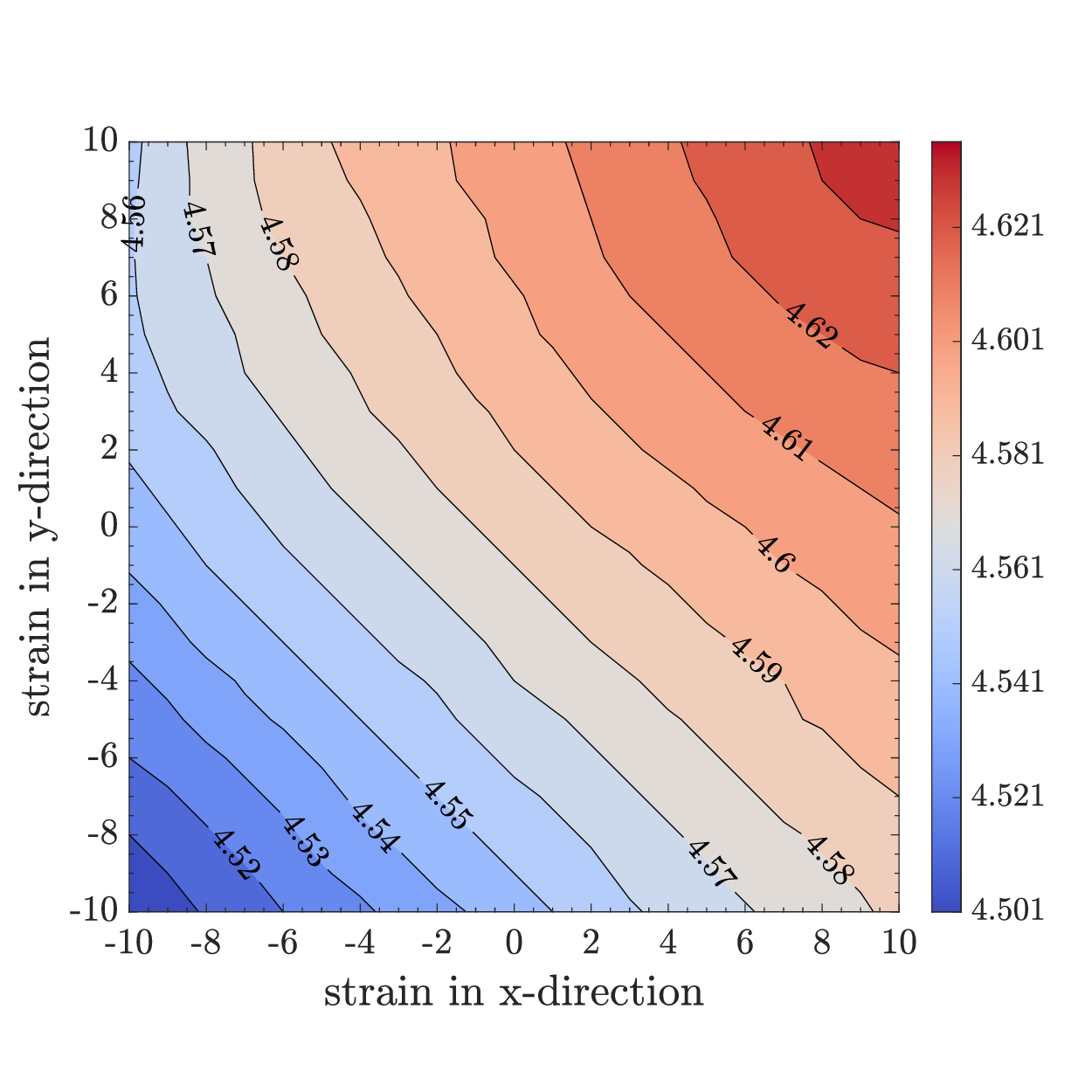}}
{\caption{Effect of strain ($\%$) on the magnetic moment of Mn.}\label{Fig:MomentvsStrain}}
\end{figure}

\paragraph{Magnetic moment - }
Figure \ref{Fig:MomentvsStrain} shows the variation of the Mn-projected local magnetic moment with applied strain. We see a clear dependence of the magnetic moment on the two-dimensional volumetric strain, $\varepsilon_{xx}+\varepsilon_{yy}$, with compressive volumetric strain reducing the local magnetic moment and tensile volumetric strain increasing it. This behavior can be understood in terms of strain-induced changes in the Mn local environment and the degree of localization of the Mn $3d$ electrons. Mn in these materials adopts a predominantly high-spin $3d^5$ configuration, with the magnetic moment arising primarily from the strongly spin-polarized Mn $3d$ states \cite{xue2020control,li2019magnetic}. Tensile volumetric strain expands the in-plane lattice and generally increases the separation between neighboring Mn atoms, thereby reducing orbital overlap and favoring greater localization of the Mn $3d$ electrons. The enhanced localization increases the spin density associated with the Mn sites and consequently increases the Mn-projected local magnetic moment. Consistent with this interpretation, recent first-principles calculations for monolayer MnBi$_2$Te$_4$ have shown that biaxial tensile strain enhances the Mn magnetic moment and attributed this behavior to the increased Mn--Mn separation and stabilization of a more localized spin state \cite{hung2025strain}. Conversely, compressive volumetric strain decreases the interatomic separations and enhances orbital overlap and Mn $3d$--(Te/Se) $p$ hybridization. This increased hybridization promotes a greater redistribution of spin density between Mn and the surrounding ligand states, resulting in a smaller magnetic moment projected onto the Mn sites. We remark that the reduction in the Mn-projected local magnetic moment does not change the Mn$^{2+}$ oxidation state or the high-spin $S=5/2$ configuration. 

The sensitivity of the magnetic properties of MnSb$_2$Te$_4$ and Mn(Bi/Sb)$_2$Se$_4$ to the changes in the Mn--Te bonding geometry is also consistent with previous first-principles studies showing that strain modifies the Mn--(Te/Se)--Mn bond angles, $p$--$d$ hybridization, and the associated magnetic interactions \cite{xue2020control,li2019magnetic}. The observed dependence of the Mn-projected local magnetic moment on $\varepsilon_{xx}+\varepsilon_{yy}$ can therefore be understood as a magnetoelastic response in which lattice expansion favors greater Mn $3d$ localization and spin polarization, whereas lattice compression enhances hybridization and redistributes part of the spin density away from the Mn sites. 

Overall, our results demonstrate a strong magnetoelastic coupling between lattice deformation, Mn--Te hybridization, and the localization of the Mn $3d$ states.

Interestingly, the strain dependence of the local magnetic moment is well described over the strain range considered by an empirical power law of the form
\begin{eqnarray}
S(\varepsilon_{xx},\varepsilon_{yy})
\approx
S_1\left(1+\varepsilon_{xx}+\varepsilon_{yy}\right)^n+S_0,
\end{eqnarray}
where $S_0$, $S_1$, and the exponent $n$ are fitting parameters given in Table \ref{Table:MomentFit}. This relation provides a compact empirical description of the calculated strain dependence over the range investigated. The good agreement with this empirical power law is consistent with our observation that the local magnetic moment responds predominantly to the two-dimensional volumetric strain, $\varepsilon_{xx}+\varepsilon_{yy}$, rather than independently to the individual strain components. 

\begin{table}[]\centering
\caption{\label{Table:MomentFit}Constants of power-law fit of magnetic moment versus strain}
\begin{tabular}{llllll}
\hline\hline
material \,\,\,\, & $S_0$  & $S_1$ & $n$ & $RMSE$ & $R^2$ \\
\hline
\hline
MnBi$_2$Se$_4$ & $4.87$ & $-0.26$ & $-1.08$ & $0.0032$ & $0.9845$  \\
MnSb$_2$Se$_4$ & $4.90$ & $-0.32$ & $-1.01$ & $0.0026$ & $0.9913$  \\
MnSb$_2$Te$_4$ & $4.99$ & $-0.42$ & $-0.78$ & $0.0032$ & $0.9874$  \\
\hline
\end{tabular}
\end{table}

\begin{figure}[h]\centering
\subfigure[MnBi$_2$Se$_4$]{\includegraphics[keepaspectratio=true,width=0.3\textwidth]{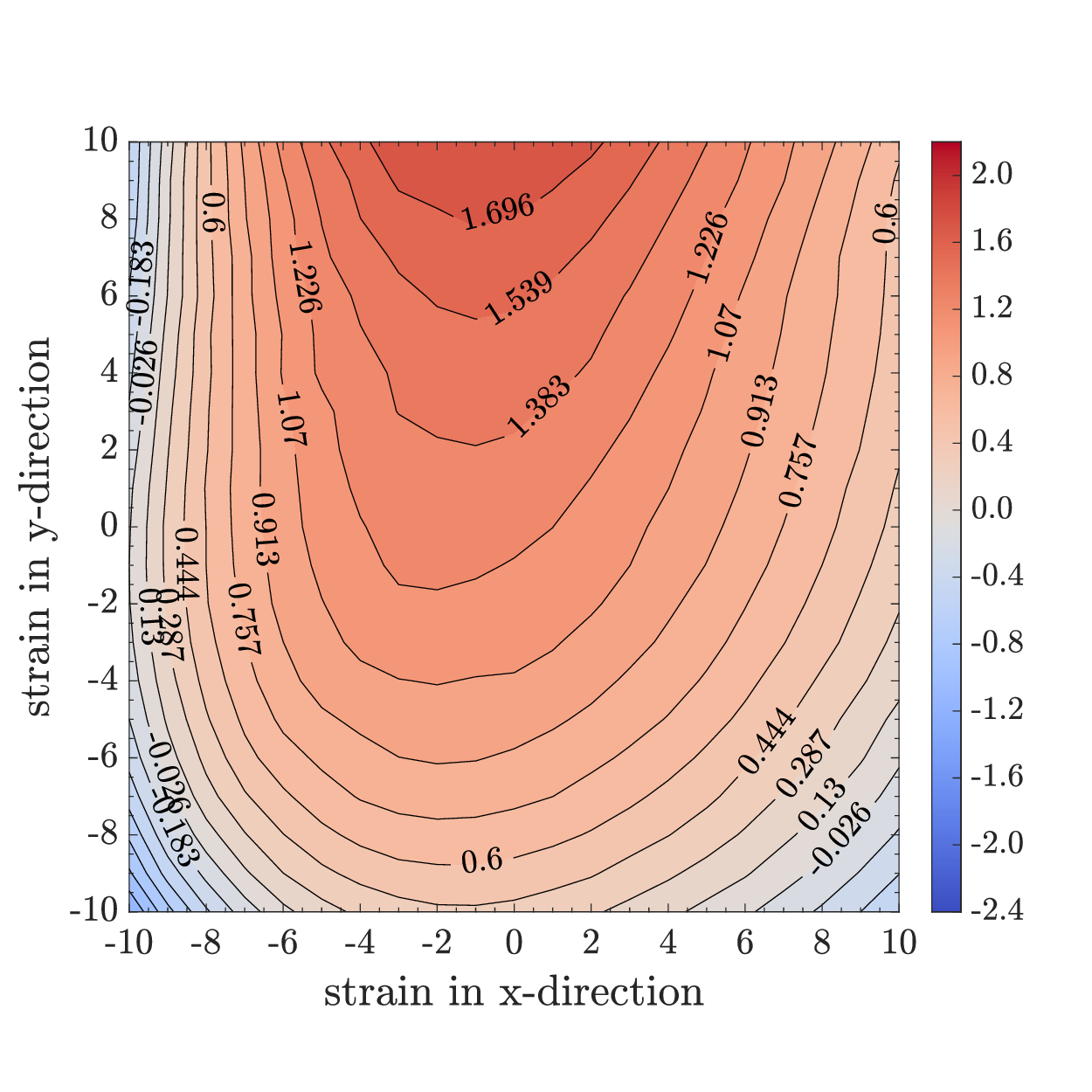}}
 \hspace{-1em}
\subfigure[MnSb$_2$Se$_4$]{\includegraphics[keepaspectratio=true,width=0.3\textwidth]{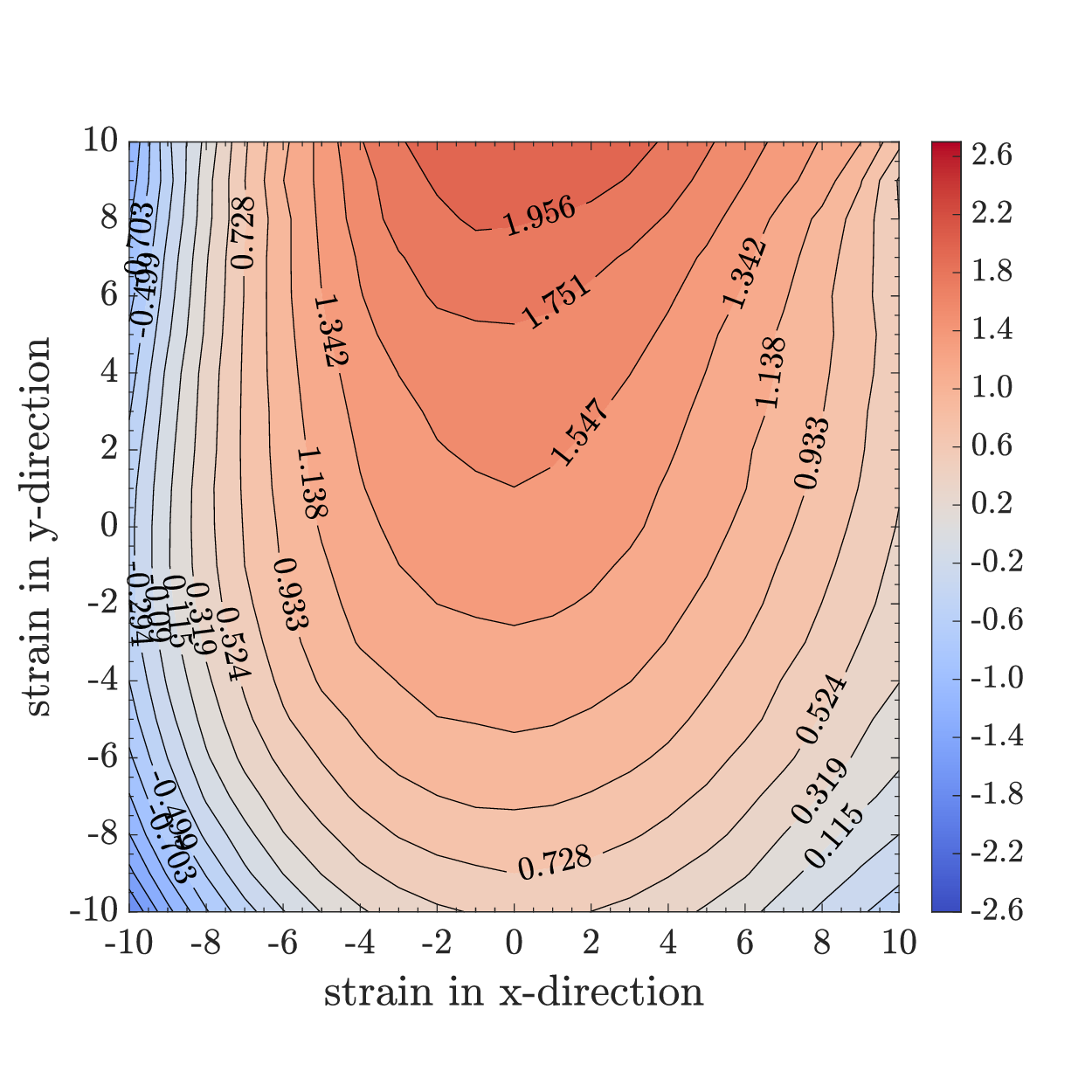}}
 \hspace{-1em}
\subfigure[MnSb$_2$Te$_4$]{\includegraphics[keepaspectratio=true,width=0.3\textwidth]{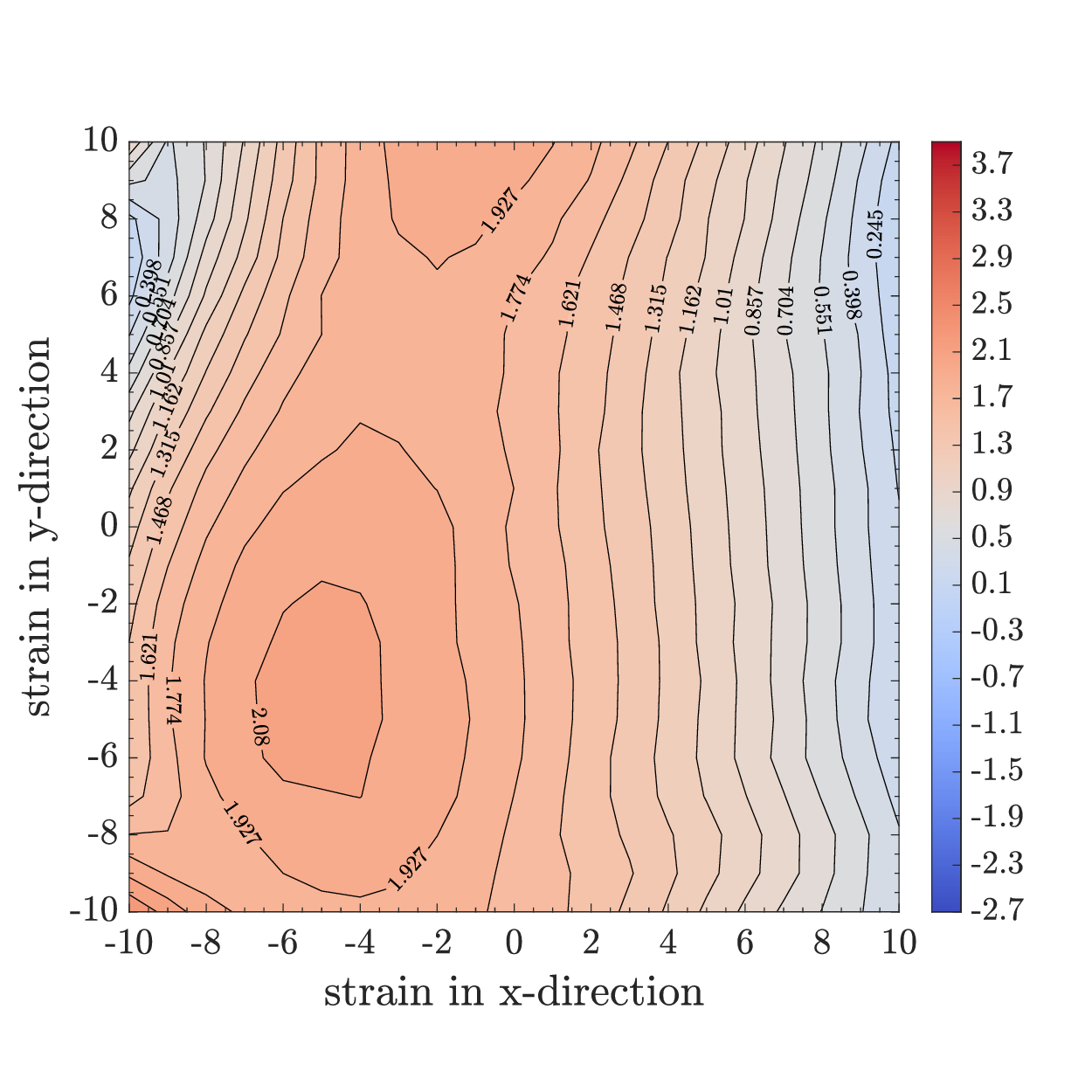}}
{\caption{Effect of strain ($\%$) on the first nearest neighbor magnetic exchange constant $J_1$ (meV).}\label{Fig:J1vsStrain}}
\end{figure}

\paragraph{Magnetic exchange - }
Figure \ref{Fig:J1vsStrain} shows the dependence of the first-nearest-neighbor exchange parameter, $J_1$, on the in-plane strain components $\varepsilon_{xx}$ and $\varepsilon_{yy}$ for MnBi$_2$Se$_4$, MnSb$_2$Se$_4$, and MnSb$_2$Te$_4$. Although all three compounds exhibit a pronounced strain dependence, the response is strongly anisotropic and, for strain along the $y$-direction, material-dependent. In MnBi$_2$Se$_4$, as $|\varepsilon_{xx}|$ becomes sufficiently large, $J_1$ crosses zero and becomes negative, indicating a change from ferromagnetic to antiferromagnetic first-nearest-neighbor coupling. The response to $\varepsilon_{yy}$ is distinctly different. In the two Se-based compounds, MnBi$_2$Se$_4$ and MnSb$_2$Se$_4$, tensile strain along the $y$-direction increases $J_1$, whereas compressive strain decreases it. In contrast, $J_1$ in MnSb$_2$Te$_4$ is comparatively insensitive to strain along the $y$-direction. Thus, while the suppression of ferromagnetic $J_1$ by $x$-directed strain is common to all three materials, the response to $y$-directed strain clearly distinguishes the Se-based compounds from MnSb$_2$Te$_4$.

The anisotropic and material-dependent behavior of $J_1$ suggests that strain modifies the detailed geometry of the exchange pathways rather than simply changing the Mn--Mn separation. The intralayer magnetic coupling in this family can be understood as arising from competition between antiferromagnetic Mn--Mn direct exchange and ligand-mediated Mn--$X$--Mn superexchange, where $X$ denotes Se or Te \cite{li2019magnetic,xu2022hydrostatic}. The direct-exchange contribution is governed by the overlap between neighboring Mn $3d$ orbitals and is therefore strongly sensitive to the Mn--Mn separation, whereas the superexchange contribution depends on the Mn--$X$--Mn bond geometry and the associated Mn $3d$--$X$ $p$ hopping and hybridization \cite{li2019magnetic,xu2022hydrostatic}. Strain along the $x$- and $y$- directions modifies these structural parameters differently, providing a microscopic basis for the pronounced directional dependence of $J_1$. Chemical species also influence the exchange pathways, and previous calculations have shown stronger Mn--Te hybridization than Mn--Se hybridization, stemming from differences in the ligand electronic structure and bonding \cite{xu2022hydrostatic}, and therefore, the weaker sensitivity of $J_1$ to $\varepsilon_{yy}$ for MnSb$_2$Te$_4$ compared with MnBi$_2$Se$_4$ and MnSb$_2$Se$_4$, as observed here.

\begin{figure}[h]\centering
\subfigure[MnBi$_2$Se$_4$ ]{\includegraphics[keepaspectratio=true,width=0.3\textwidth]{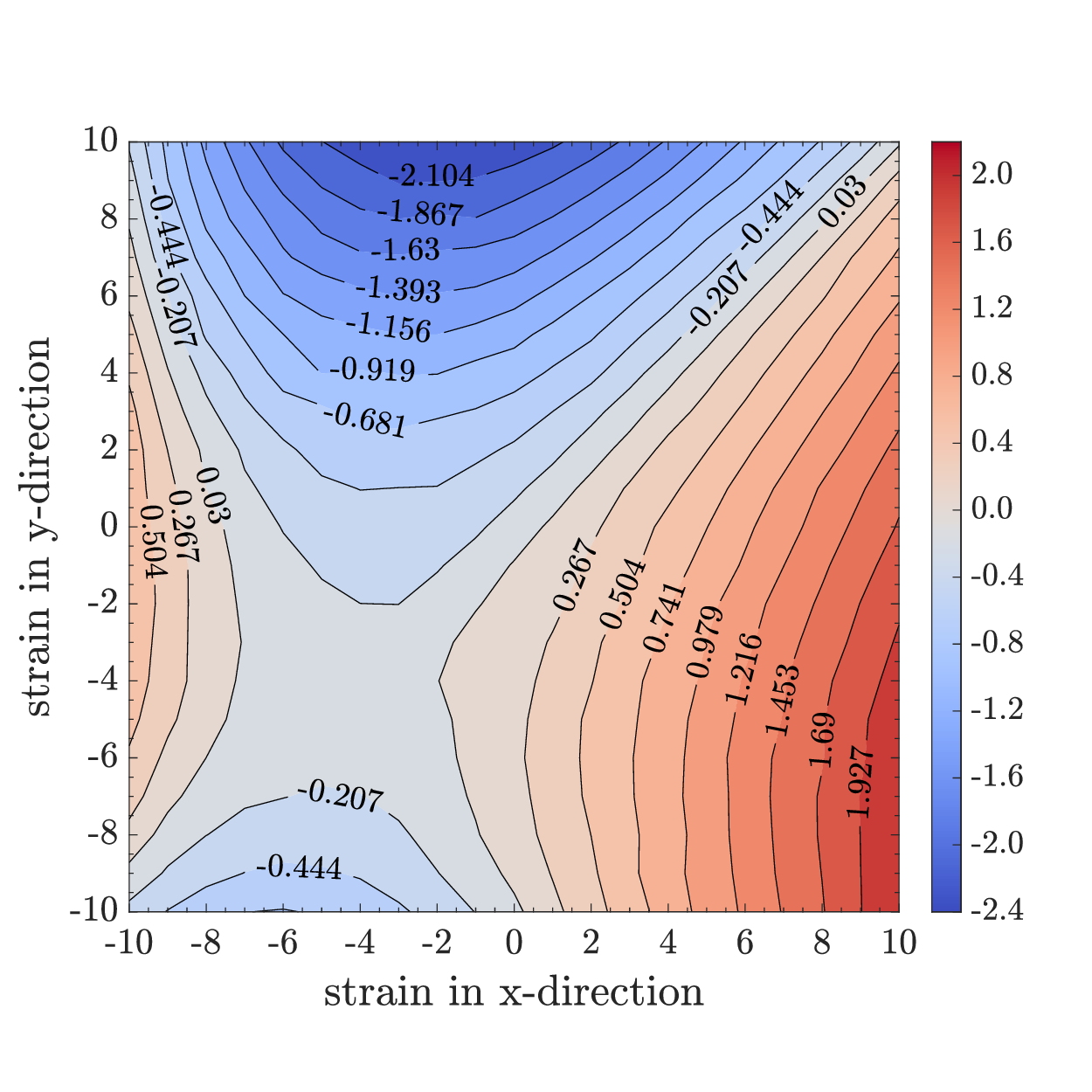}}
 \hspace{-1em}
\subfigure[MnSb$_2$Se$_4$]{\includegraphics[keepaspectratio=true,width=0.3\textwidth]{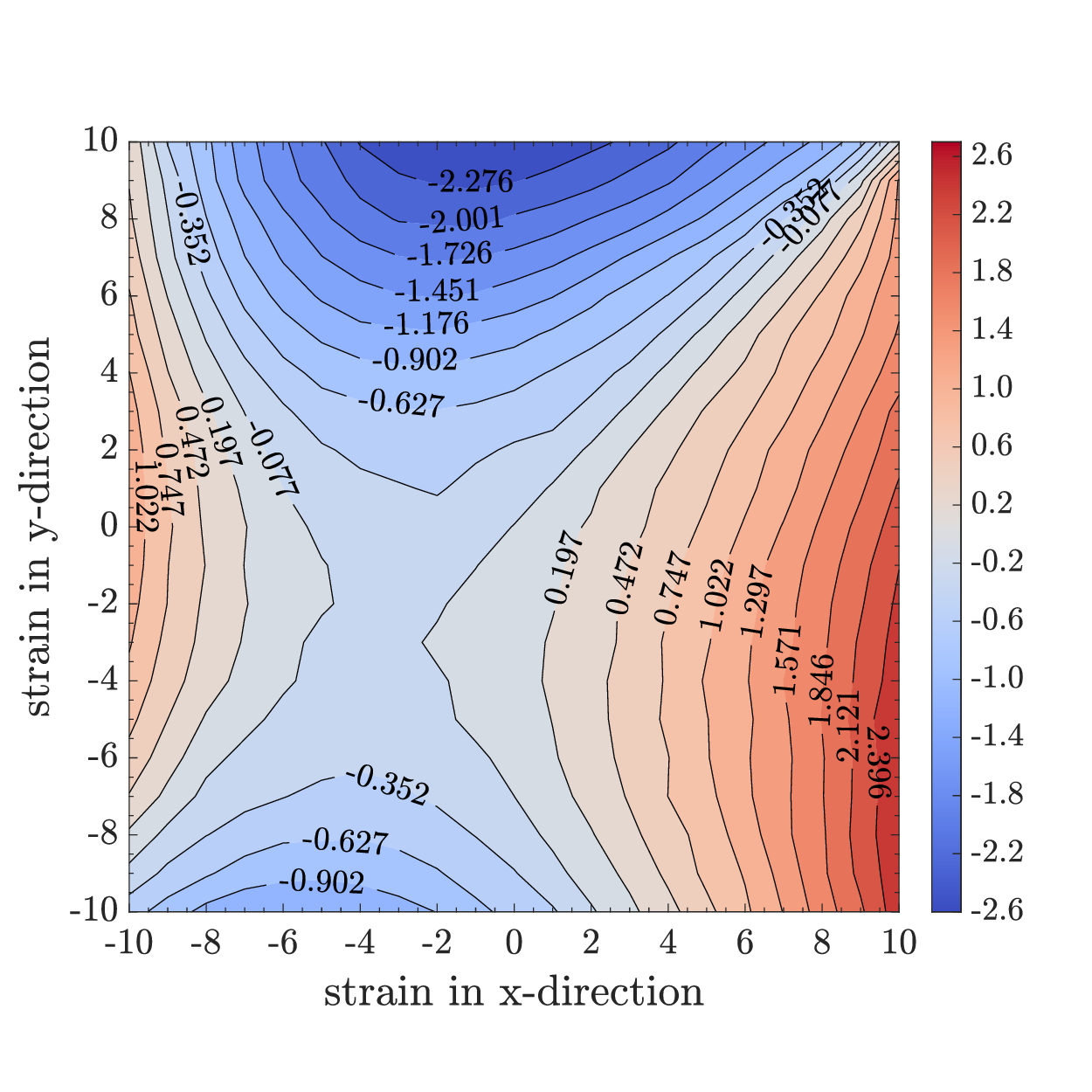}}
 \hspace{-1em}
\subfigure[MnSb$_2$Te$_4$ ]{\includegraphics[keepaspectratio=true,width=0.3\textwidth]{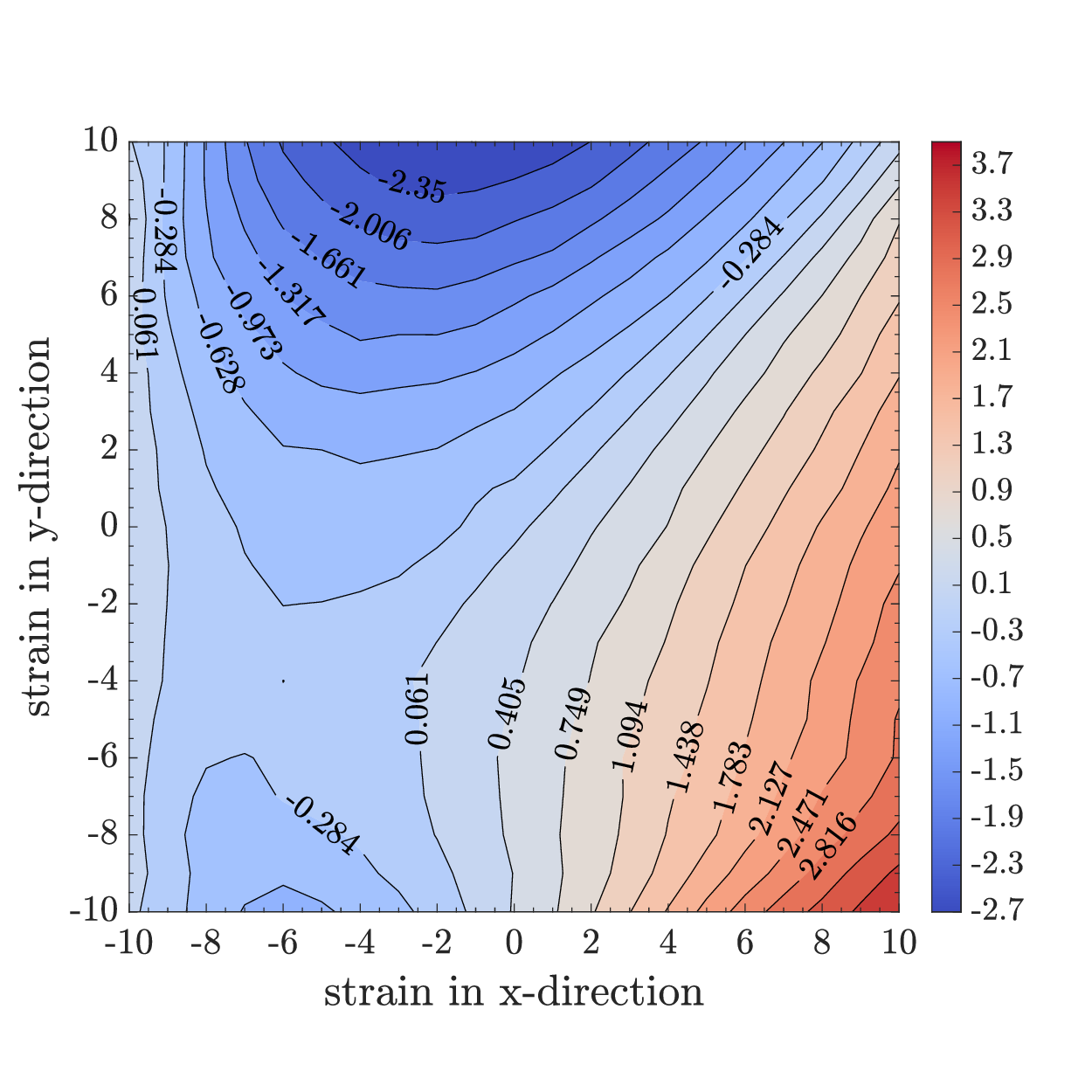}}
{\caption{Effect of strain ($\%$) on the second nearest neighbor magnetic exchange constant $J_2$ (meV).}\label{Fig:J2vsStrain}}
\end{figure}

\begin{figure}[h]\centering
\subfigure[MnBi$_2$Se$_4$]{\includegraphics[keepaspectratio=true,width=0.3\textwidth]{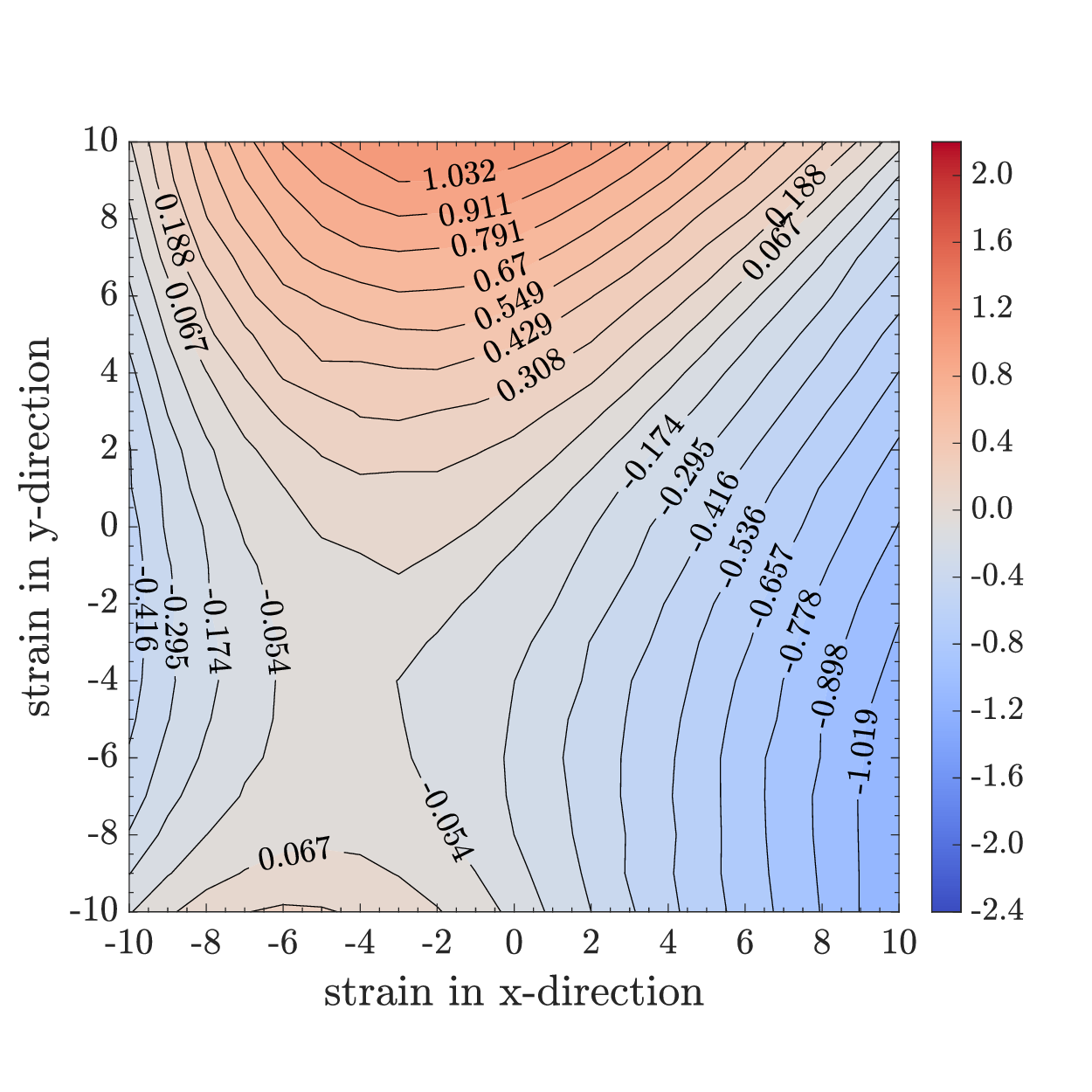}} 
 \hspace{-1em}
\subfigure[MnSb$_2$Se$_4$]{\includegraphics[keepaspectratio=true,width=0.3\textwidth]{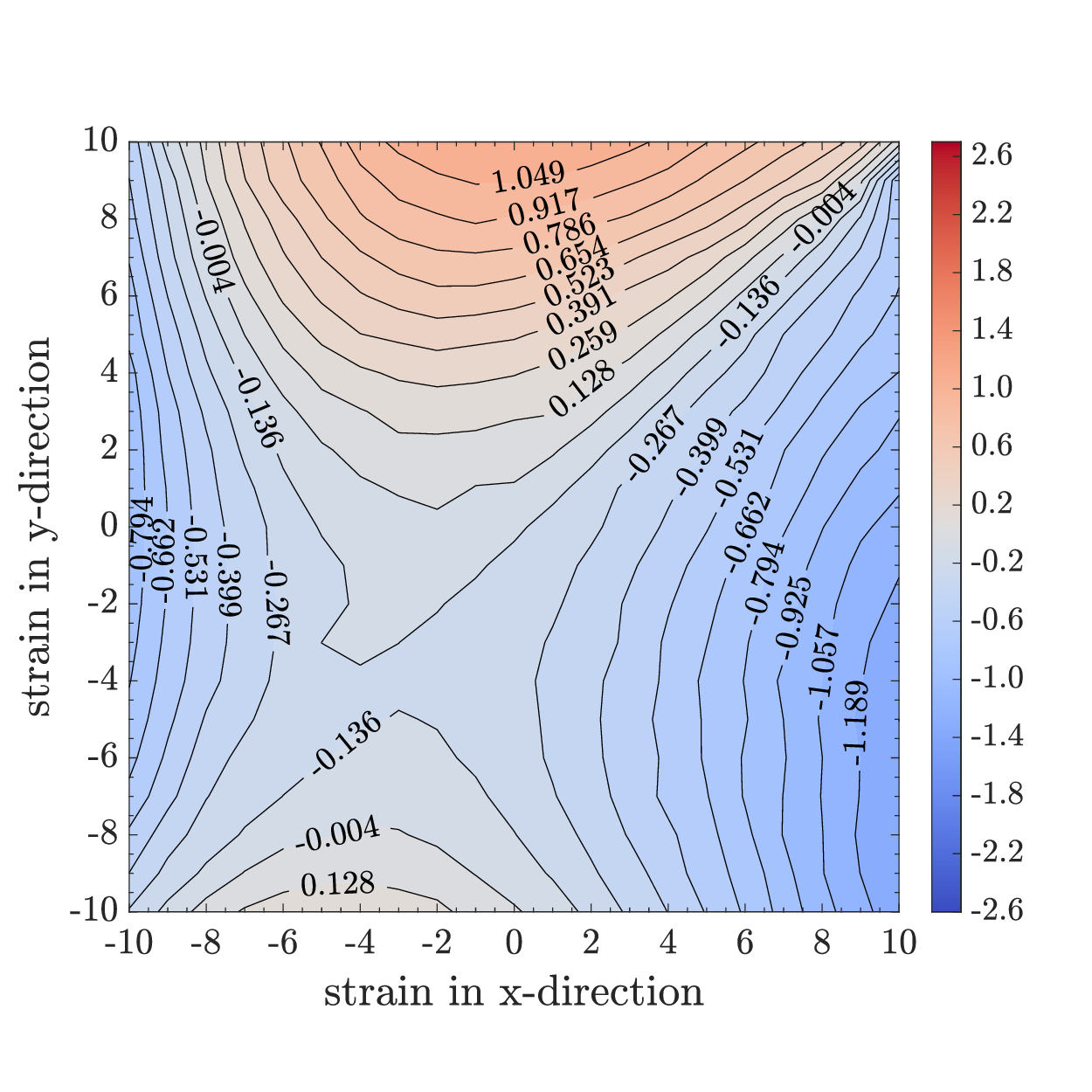}} 
 \hspace{-1em}
\subfigure[MnSb$_2$Te$_4$]{\includegraphics[keepaspectratio=true,width=0.3\textwidth]{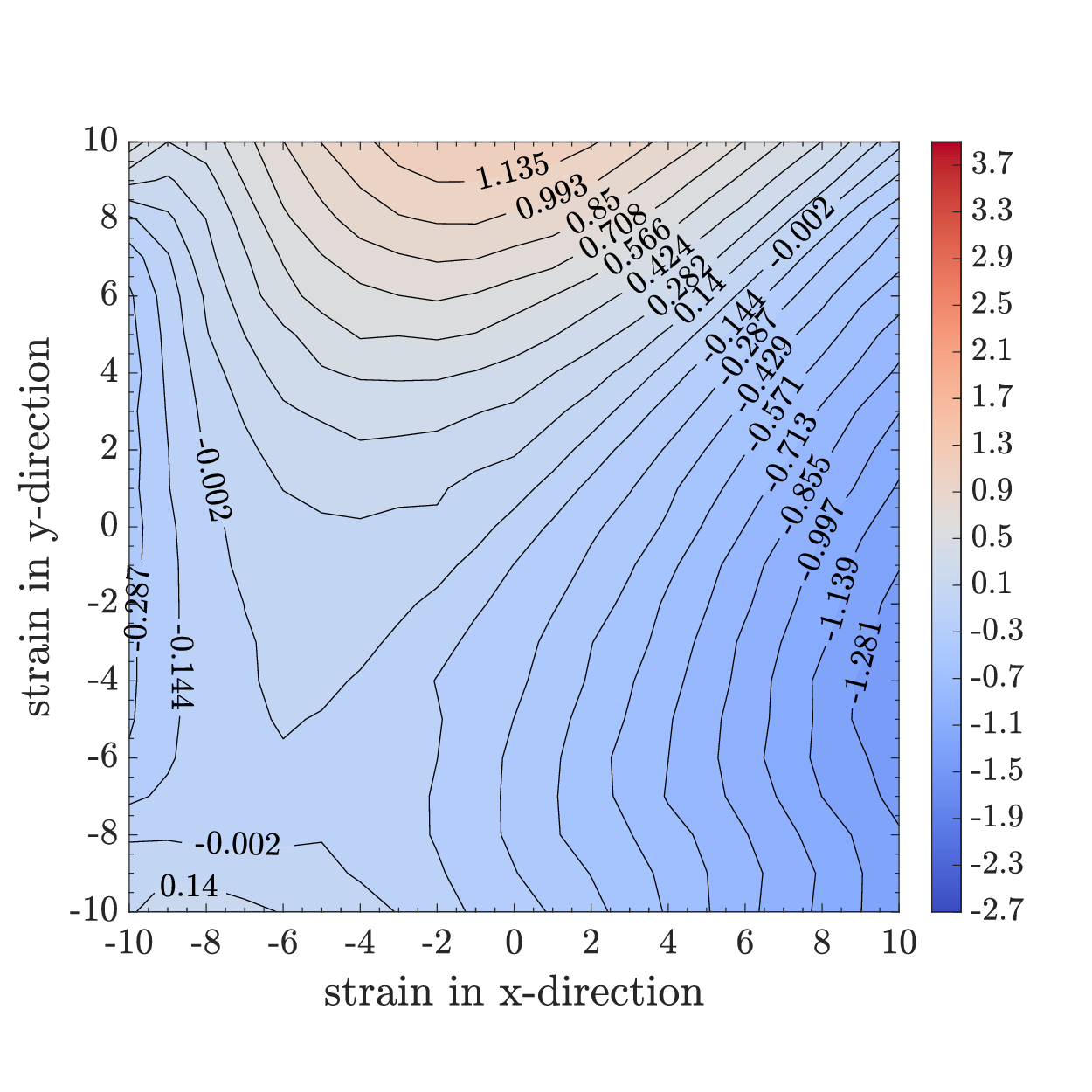}} 
{\caption{Effect of strain ($\%$) on the third nearest neighbor magnetic exchange constant $J_3$ (meV).}\label{Fig:J3vsStrain}}
\end{figure}

Figure \ref{Fig:J2vsStrain} shows the corresponding dependence of the second-nearest-neighbor exchange parameter, $J_2$, on in-plane strain. For the three materials, increasing $\varepsilon_{xx}$ increases $J_2$, whereas increasing $\varepsilon_{yy}$ decreases it. The third-nearest-neighbor exchange parameter, $J_3$, exhibits the opposite directional response where increasing $\varepsilon_{xx}$ decreases $J_3$, whereas increasing $\varepsilon_{yy}$ increases it, as shown in Figure \ref{Fig:J3vsStrain}. The opposing strain dependences of $J_2$ and $J_3$ indicate that these interactions are associated with geometrically distinct exchange pathways that respond differently to the same lattice deformation. Consequently, strain can strengthen one exchange interaction while simultaneously weakening another, thereby modifying not only the overall strength of magnetic coupling but also the competition among first-, second-, and third-nearest-neighbor interactions.

Overall, these results demonstrate a strongly anisotropic and material-dependent magnetoelastic coupling in MnBi$_2$Se$_4$, MnSb$_2$Se$_4$, and MnSb$_2$Te$_4$. In particular, the strain-induced sign reversal of $J_1$ shows that sufficiently large in-plane deformation can alter the balance between competing ferromagnetic and antiferromagnetic exchange mechanisms rather than simply renormalizing the magnitude of an otherwise unchanged interaction. Since $J_1$ is the dominant exchange interaction in these materials, its reversal from positive to negative provides a microscopic mechanism for the emergence of antiferromagnetic ordering in the corresponding regions of the magnetic phase diagrams shown in Figure \ref{PhaseDiagram}. At the same time, the contrasting responses of $J_2$ and $J_3$ show that the resulting magnetic phase cannot be understood from $J_1$ alone, as strain simultaneously modifies the longer-range exchange interactions and can therefore change the degree of competition among different magnetic configurations.

The directional dependence of the magnetic response can also be understood from the symmetry breaking induced by anisotropic in-plane strain. In the unstrained state, the monolayers crystallize in the trigonal $P\bar{3}m1$ space group and possess threefold rotational $C_{3z}$ symmetry about the out-of-plane axis. This symmetry renders the symmetry-related in-plane Mn--Mn exchange pathways equivalent. Equibiaxial strain preserves the $C_{3z}$ symmetry, whereas a general in-plane strain state with $\varepsilon_{xx}\neq\varepsilon_{yy}$ lowers the crystal symmetry and removes the equivalence between exchange pathways oriented along different in-plane directions. As a result, strain applied along the $x$- and $y$-directions modifies the Mn--Mn separations, Mn--$X$ bond lengths, Mn--$X$--Mn bond angles, and Mn $3d$--$X$ $p$ hybridization differently, producing an anisotropic magnetoelastic response even when the exchange interaction is approximately isotropic in spin space. Notably, the symmetry breaking under anisotropic strain splits a given neighbor shell into non-equivalent bond-specific exchange interactions; therefore, the $J_1$, $J_2$, and $J_3$ values extracted here from the four-state energy mapping should be interpreted as effective exchange parameters for the corresponding neighbor shells. The DFT magnetic phase diagrams, by contrast, do not rely on this effective Heisenberg representation and are obtained directly from the total-energy differences among the competing magnetic configurations.

\paragraph{Magnetocrystalline anisotropy - }
The magnetocrystalline anisotropy (MCA), $\eta$, was evaluated using the energy difference between two magnetization directions. First, a self-consistent collinear calculation without spin--orbit coupling (SOC) was performed with the $ 8$-atom primitive unit cell to obtain the converged ground-state charge density. This charge density was then used in two non-self-consistent SOC calculations with the spin quantization axis oriented along the in-plane $x$ direction and the out-of-plane $z$ direction, respectively \cite{xue2020control}. The MCA was calculated as

\begin{equation}
\eta = E_x-E_z ,
\end{equation}

where $E_x$ and $E_z$ are the total energies for magnetization along the $x$ and $z$ directions, respectively. With this convention, $\eta>0$ indicates that the $z$ direction is energetically preferred, corresponding to an out-of-plane easy axis, whereas $\eta<0$ indicates a preference for in-plane magnetization.

At zero strain, the calculated MCA values are $1.16$ meV for MnBi$_2$Se$_4$, $1.17$ meV for MnSb$_2$Se$_4$, and $1.27$ meV for MnSb$_2$Te$_4$. Thus, all three monolayers favor out-of-plane magnetization in their unstrained configurations.

\begin{figure}[h]\centering
\subfigure[MnBi$_2$Se$_4$ ]{\includegraphics[keepaspectratio=true,width=0.3\textwidth]{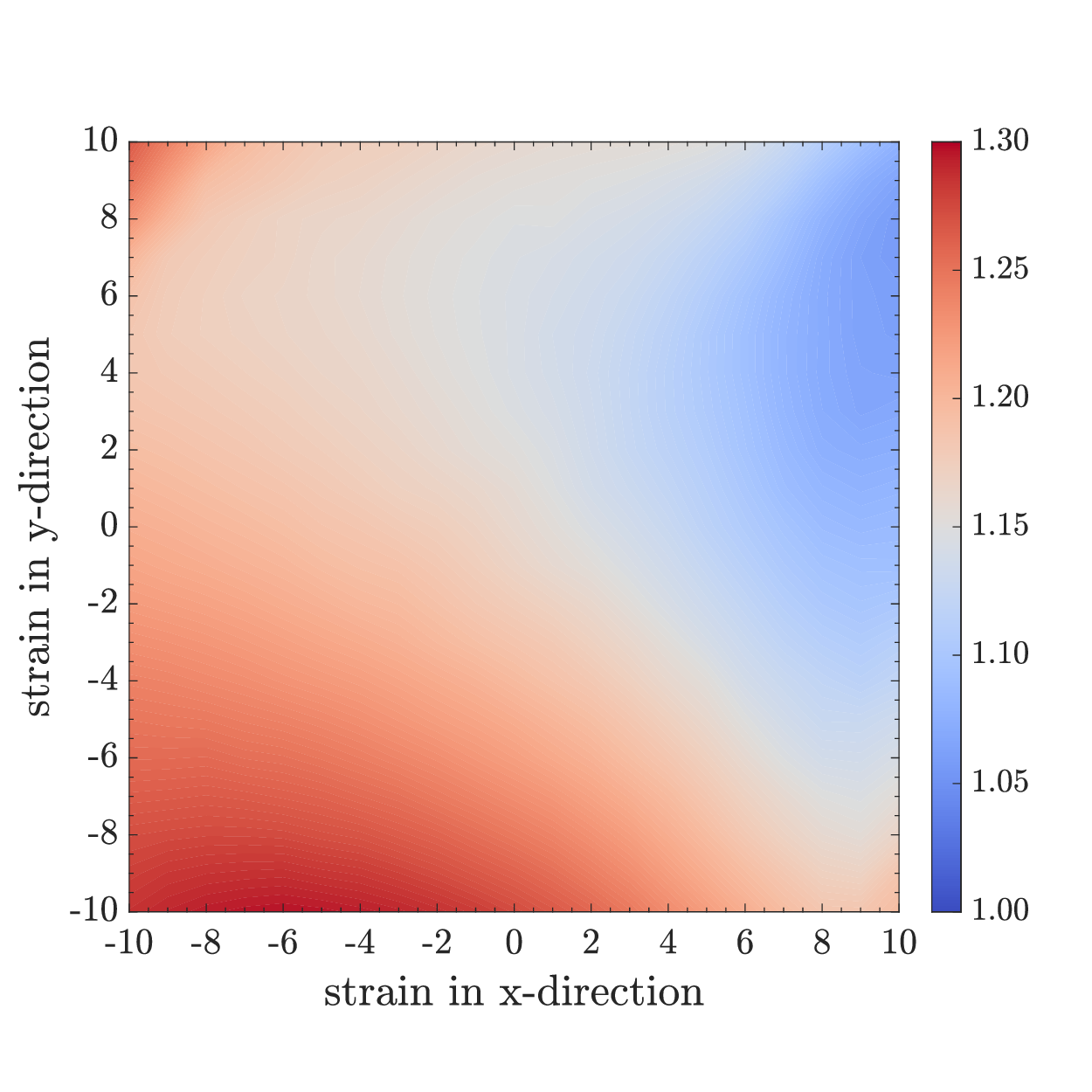}} 
\subfigure[MnSb$_2$Se$_4$ ]{\includegraphics[keepaspectratio=true,width=0.3\textwidth]{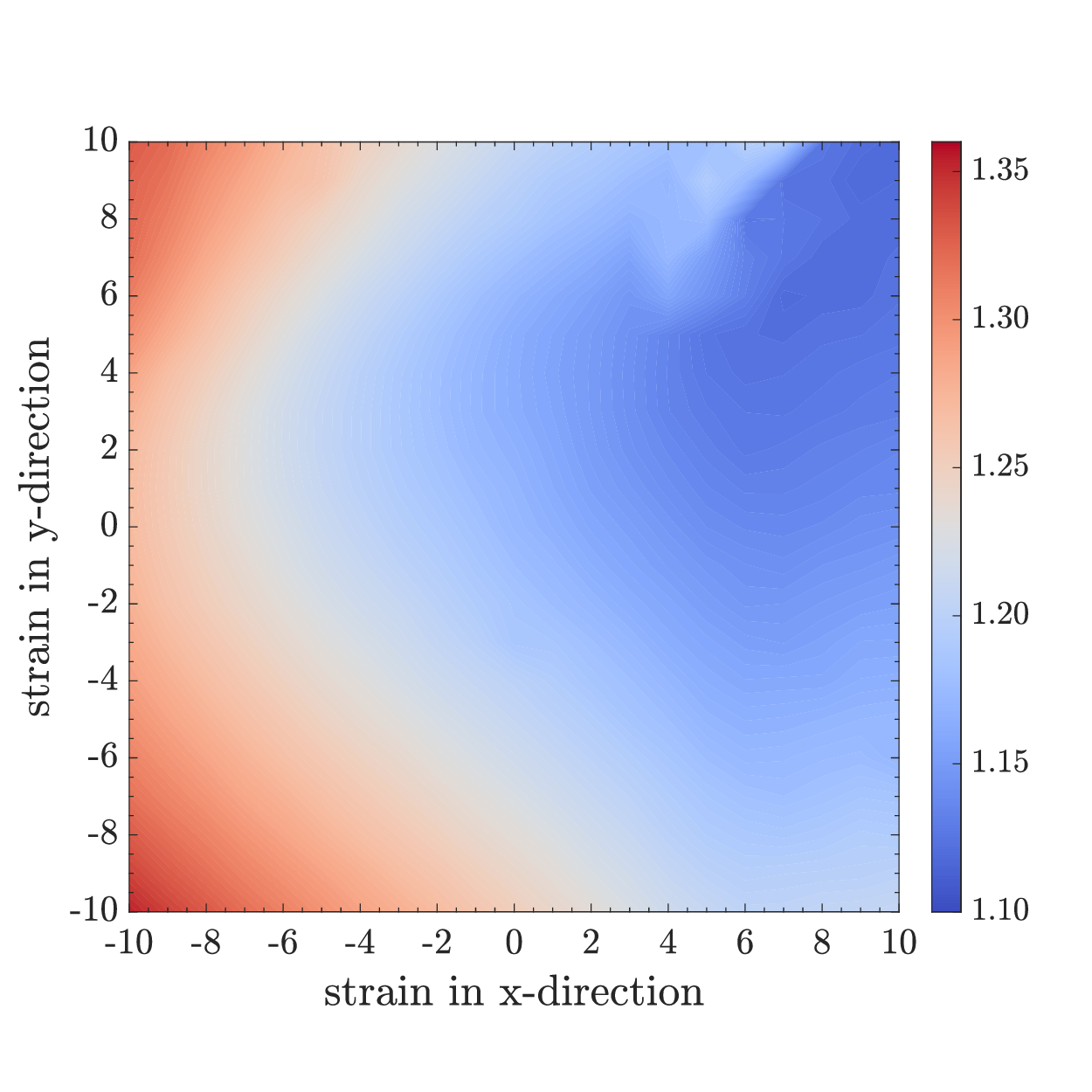}} 
\subfigure[MnSb$_2$Te$_4$ ]{\includegraphics[keepaspectratio=true,width=0.3\textwidth]{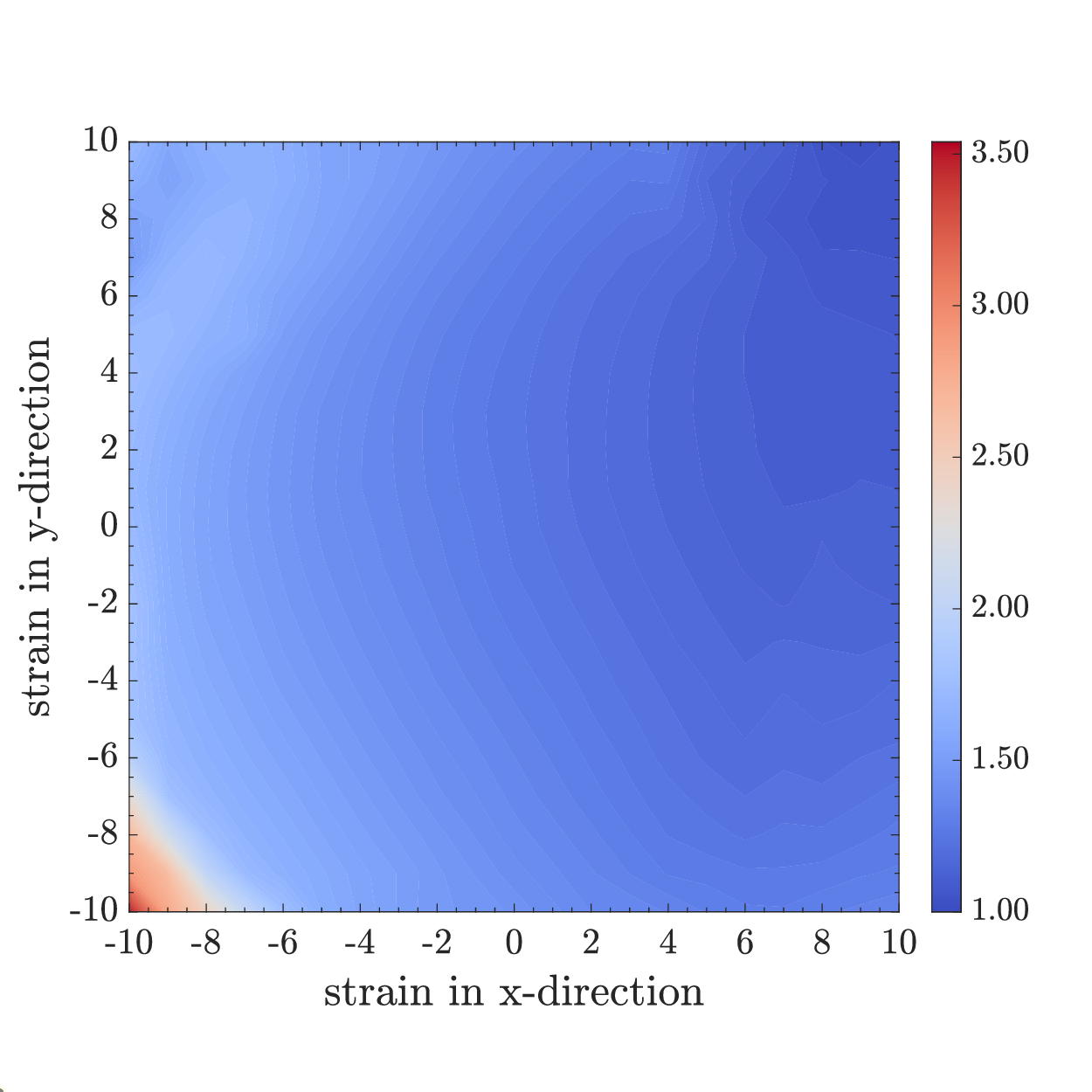}} 
{\caption{Effect of strain ($\%$) on the magnetic anisotropy $\eta$ (meV)}\label{Fig:anisotropyvsStrain}}
\end{figure}

The strain dependence of the MCA is shown in Figure \ref{Fig:anisotropyvsStrain}. For MnBi$_2$Se$_4$ and MnSb$_2$Se$_4$, compressive strain enhances $\eta$, whereas tensile strain reduces it, demonstrating a finite magnetoelastic coupling between the lattice deformation and the magnetic anisotropy. The larger magnitude of MCA of MnSb$_2$Te$_4$ can be understood in the context of the microscopic origin of the magnetic anisotropy in this family. Previous first-principles calculations \cite{li2019magnetic} have shown that the magnetic anisotropy of the Te-based MnBi$_2$Te$_4$-type monolayers is dominated by single-ion anisotropy arising from the combined spin--orbit coupling of Mn and Te. In particular, the calculated single-ion anisotropy of MnSb$_2$Te$_4$ is comparable to that of MnBi$_2$Te$_4$, whereas replacing Te with Se reduces this contribution because of the weaker spin--orbit coupling of Se. As a result, exchange anisotropy becomes more important in the Se-based compounds. The strong Te-assisted single-ion anisotropy therefore provides a microscopic origin for the larger zero-strain MCA of MnSb$_2$Te$_4$. In the Se-based compounds, where the single-ion anisotropy is weaker, strain-induced modifications of the crystal-field environment, $p$--$d$ hybridization, and exchange anisotropy constitute a larger fraction of the total anisotropy, resulting in a more pronounced strain dependence.

The MCA of MnSb$_2$Te$_4$ remains remarkably insensitive to strain over the range considered, despite its larger magnitude. However, this weak strain dependence of the MCA in MnSb$_2$Te$_4$ does not imply that its electronic structure is insensitive to strain, since previous calculations have demonstrated strain-induced modifications of its electronic bands and topological properties \cite{li2021electronic}. Rather, the results indicate that the net SOC-induced energy difference between the two magnetization directions remains comparatively stable. Band-resolved calculations for the related MnBi$_2$Te$_4$ system have further shown that the total MCA contains electronic contributions of opposite sign \cite{xue2020control}, suggesting that partial compensation between strain-dependent contributions can also contribute to the weak variation observed here.

\paragraph{Magnetic ordering temperature - }
We calculate the magnetic ordering temperature by first fitting a classical Heisenberg Hamiltonian
\begin{eqnarray}\label{Eq:HamiltonianWithAnisotropy}
    H = -\frac{1}{2} \sum_{i,j} J_{ij} \,\bm{S}_i \cdot \bm{S}_j -\eta \sum_i (\bm{S}_i\cdot \hat{\bm{e}}_z)^2 \,\,,
\end{eqnarray}
and we use this Hamiltonian to perform automated high throughput Metropolis Monte-Carlo simulations using a $16\times 16$ two-dimensional hexagonal lattice, scanning temperature and strain.  For each system, the temperature was varied from $1$ Kelvin to $40$ Kelvin using $0.2$ Kelvin intervals, and the specific heat $c_v$ was calculated using equilibrium energy fluctuations $c_v = (\langle E^2\rangle - \langle E\rangle^2)/(k_BT)^2$. We identify the phase transition temperature with the maximum of $c_v$ for this finite-size simulation.

\begin{figure}[h]\centering
\subfigure[MnBi$_2$Se$_4$]{\includegraphics[keepaspectratio=true,width=0.32\textwidth]{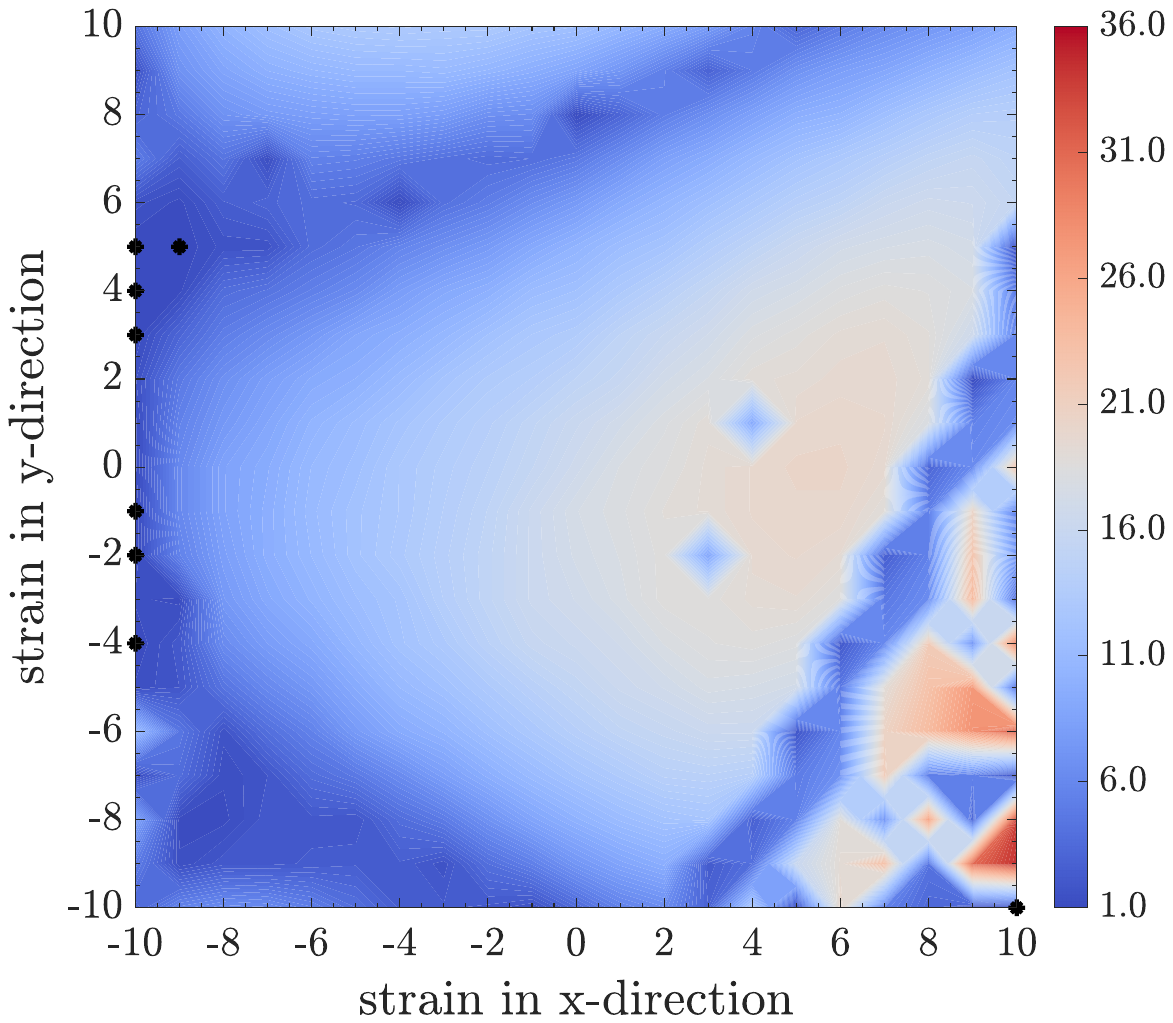}} 
\subfigure[MnSb$_2$Se$_4$]{\includegraphics[keepaspectratio=true,width=0.32\textwidth]{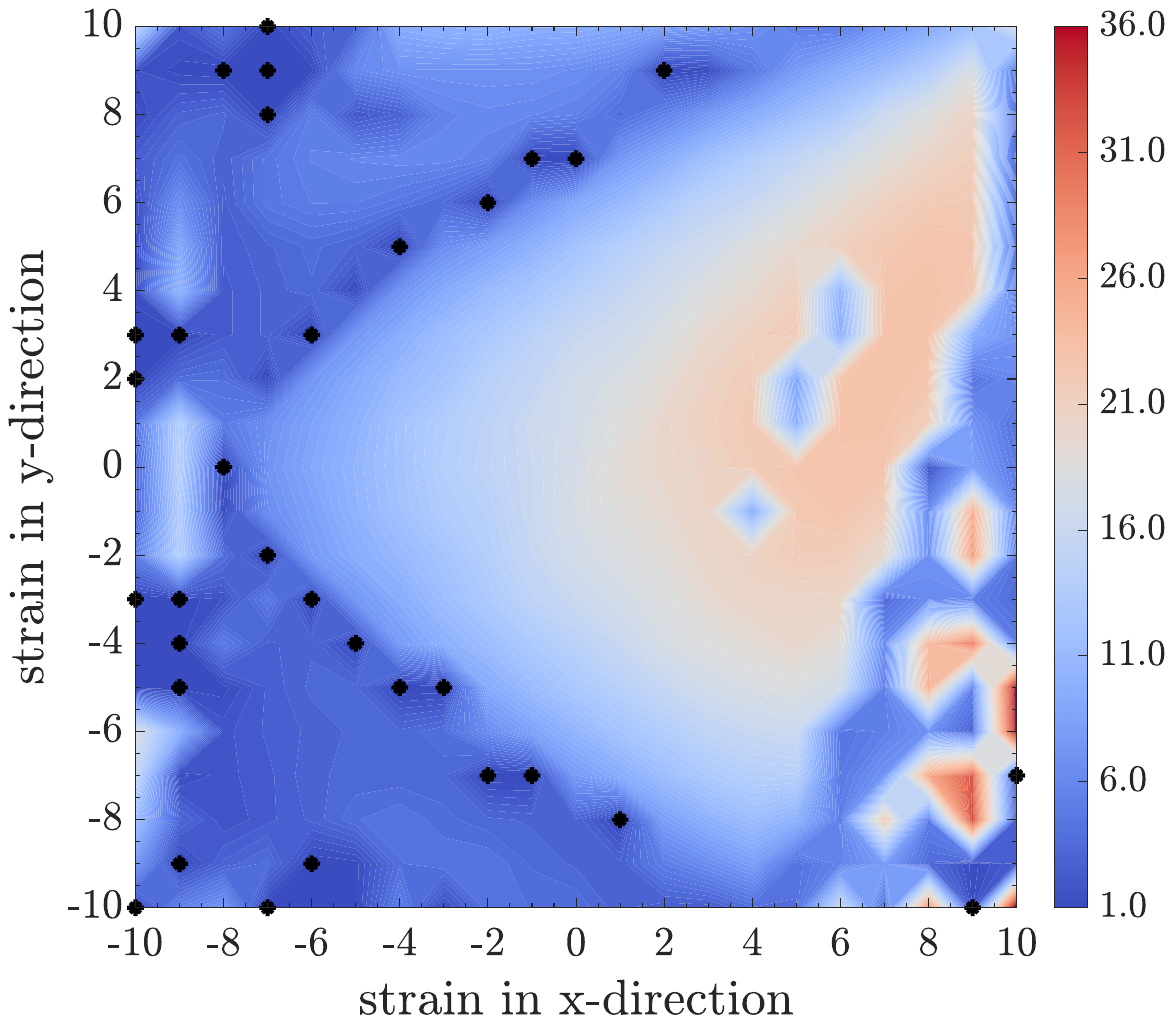}} 
\subfigure[MnSb$_2$Te$_4$]{\includegraphics[keepaspectratio=true,width=0.32\textwidth]{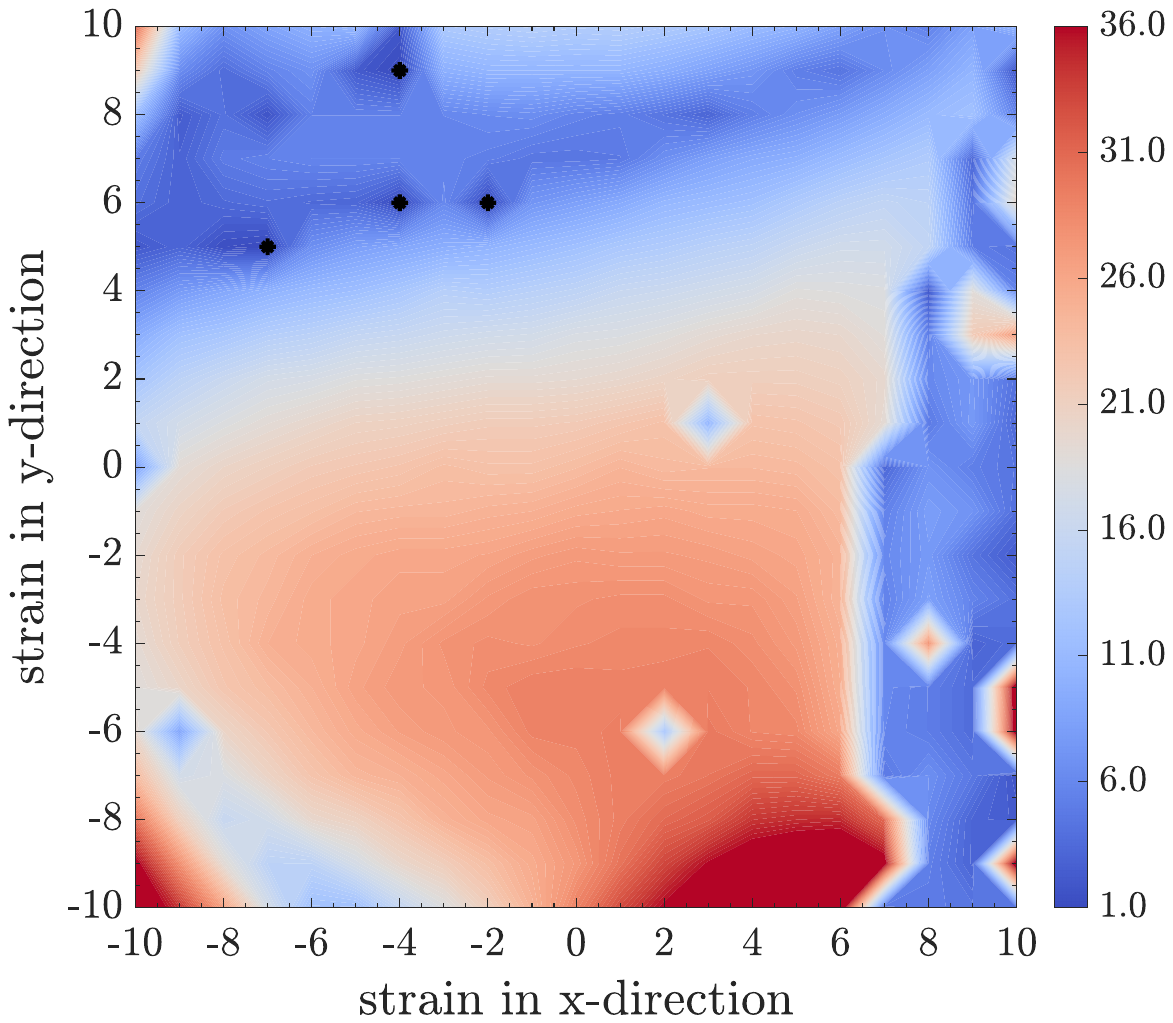}} 
{\caption{The magnetic ordering temperature $T_{\mathrm{C}}$ vs strain ($\%$). The unit of temperature is Kelvin. The black dots are points where the Monte Carlo simulations did not converge.}\label{Fig:TcvsStrain}}
\end{figure}

Figure~\ref{Fig:TcvsStrain} shows the dependence of the magnetic ordering temperature, $T_{\mathrm{C}}$, of MnBi$_2$Se$_4$, MnSb$_2$Se$_4$, and MnSb$_2$Te$_4$ on the two in-plane strain components, $\varepsilon_{xx}$ and $\varepsilon_{yy}$. In the absence of strain, the magnetic ordering temperatures of MnBi$_2$Se$_4$, MnSb$_2$Se$_4$, and MnSb$_2$Te$_4$ monolayers are $16.8$ Kelvin, $17.6$ Kelvin, and $24.0$ Kelvin, respectively. Note that for the tensile $x$-strain or compressive $y$-strain quadrant, these calculations showed significantly larger noise, thus leading to the artifacts in the transition temperatures seen in Figure~\ref{Fig:TcvsStrain}. The black markers in Figure~\ref{Fig:TcvsStrain} denote Monte Carlo calculations that did not converge and should therefore be regarded as missing data rather than as zero-$T_{\mathrm{C}}$ states or magnetic phase boundaries. For all three monolayers, $T_{\mathrm{C}}$ exhibits a pronounced nonlinear and anisotropic strain dependence, demonstrating that the magnetic ordering temperature is determined by the specific strain state rather than by the strain magnitude alone. This behavior is consistent with the strongly anisotropic strain dependence of the exchange parameters discussed above. Because $T_{\mathrm{C}}$ depends on the collective balance among $J_1$, $J_2$, and $J_3$, as well as on the magnetic anisotropy that stabilizes long-range magnetic order in two dimensions, changes in the individual interactions can cooperate or compete with one another as the strain state is varied. The resulting $T_{\mathrm{C}}$ landscapes therefore reflect the combined magnetoelastic response of the underlying exchange network rather than the evolution of any single exchange parameter.

For MnBi$_2$Se$_4$, as shown in Figure ~\ref{Fig:TcvsStrain}(a), compressive $\varepsilon_{xx}$ decreases $T_{\mathrm{C}}$, whereas moderate tensile $\varepsilon_{xx}$ produces a broad and smooth region of increased magnetic ordering temperature. Larger values of $T_{\mathrm{C}}$ are obtained when tensile $\varepsilon_{xx}$ is combined with compressive $\varepsilon_{yy}$. The highest values occur near the lower-right boundary of the sampled strain space and are spatially localized. Notably, the increase in $T_{\mathrm{C}}$ under tensile $\varepsilon_{xx}$ occurs even though $J_1$ does not exhibit a corresponding enhancement, indicating that the ordering temperature is not controlled by $J_1$ alone. Instead, strain-induced changes in $J_2$, $J_3$, and the magnetic anisotropy can modify the effective stability of the ferromagnetic state and thereby compensate for, or outweigh, changes in the nearest-neighbor interaction.

The effect of strain on the magnetic ordering temperature of MnSb$_2$Se$_4$ is shown in Figure.~\ref{Fig:TcvsStrain}(b). Similar to MnBi$_2$Se$_4$, the ordering temperature in MnSb$_2$Se$_4$ is also increased by tensile $\varepsilon_{xx}$; however, the region of increased $T_{\mathrm{C}}$ is shifted toward larger positive $\varepsilon_{xx}$ and extends to a higher nominal temperature range. At large tensile $\varepsilon_{xx}$, however, the $T_{\mathrm{C}}$ landscape becomes increasingly nonuniform, with neighbouring regions of high and low ordering temperature. This increased structure suggests stronger competition among the strain-dependent exchange interactions in MnSb$_2$Se$_4$, such that relatively small changes in the deformation state can significantly alter the balance among $J_1$, $J_2$, $J_3$, and $\eta$. Thus, replacing Bi with Sb in the Se-based monolayer increases the attainable strain-induced enhancement of $T_{\mathrm{C}}$, but also produces a more strain-sensitive magnetic response at large deformation.

MnSb$_2$Te$_4$, shown in Figure~\ref{Fig:TcvsStrain}(c), exhibits a qualitatively different strain landscape. The high-$T_{\mathrm{C}}$ region extends across compressive $\varepsilon_{yy}$ and over a wide range of $\varepsilon_{xx}$, with the largest converged values obtained for compressive $y$-strain combined with small-to-moderate tensile $x$-strain. Overall, $T_{\mathrm{C}}$ is considerably more sensitive to $\varepsilon_{yy}$ than to $\varepsilon_{xx}$. It is worthwhile to note that $J_1$ in MnSb$_2$Te$_4$ is comparatively insensitive to strain along the $y$ direction (see Figure \ref{Fig:J1vsStrain}(c)). The strong $\varepsilon_{yy}$ dependence of $T_{\mathrm{C}}$ therefore provides further evidence that the ordering temperature cannot be understood from the nearest-neighbor interaction alone. Instead, the opposing strain responses of $J_2$ and $J_3$ play an important role in controlling the magnetic energy scale under $y$-directed strain. Moreover, because the magnetocrystalline anisotropy of MnSb$_2$Te$_4$ remains comparatively stable over the investigated strain range (see Figure \ref{Fig:anisotropyvsStrain}c), the pronounced variation of $T_{\mathrm{C}}$ with $\varepsilon_{yy}$ is attributed more to the redistribution of the exchange interactions than to large changes in the anisotropy energy.

Within the investigated strain range, MnSb$_2$Te$_4$ exhibits the most extensive converged region of elevated $T_{\mathrm{C}}$, indicating a comparatively large strain window over which enhanced magnetic ordering can be maintained. This behavior contrasts with the more localized regions of enhanced $T_{\mathrm{C}}$ in the two selenides. The distinction between the selenides and the telluride further highlights the role of chemical species in determining how lattice deformation modifies the competing magnetic interactions.

Consequently, broad and continuously sampled regions of enhanced $T_{\mathrm{C}}$ provide a more reliable basis for identifying strain-tunable magnetic regimes than isolated extrema or narrow contour features adjacent to non-converged points. Overall, tensile $\varepsilon_{xx}$ provides the dominant route for enhancing $T_{\mathrm{C}}$ in MnBi$_2$Se$_4$ and MnSb$_2$Se$_4$, whereas compressive $\varepsilon_{yy}$ plays the dominant role in MnSb$_2$Te$_4$. The contrasting responses emphasize that strain engineering modifies magnetic interactions in a composition-dependent manner.

Overall, these results demonstrate that the magnetic ordering temperature is governed by a strongly anisotropic and chemical species-dependent magnetoelastic response. Importantly, the strain dependence of $T_{\mathrm{C}}$ does not simply track the behavior of the dominant nearest-neighbor exchange interaction, $J_1$, and instead is due to a combined evolution and competition of $J_1$, $J_2$, and $J_3$, together with the magnetocrystalline anisotropy that stabilizes long-range magnetic order in two dimensions. Strain engineering can consequently enhance the ordering temperature even when the dominant nearest-neighbor interaction is weakened, provided that the accompanying changes in the longer-range exchange interactions and magnetic anisotropy increase the overall stability of the ferromagnetic state.

\section{Concluding remarks and outlook}
In this work, we have systematically investigated the influence of in-plane strain on the magnetic properties of monolayer MnBi$_2$Se$_4$, MnSb$_2$Se$_4$, and MnSb$_2$Te$_4$. Our results demonstrate that strain provides an effective means of tuning both the local and collective magnetic behavior of these materials. The Mn-projected local magnetic moments exhibit a clear magnetoelastic response, with tensile volumetric strain generally promoting greater localization and spin polarization of the Mn $3d$ states, whereas compression enhances orbital overlap and Mn--chalcogen hybridization. More importantly, the magnetic response cannot be described by the strain magnitude alone, and the individual strain components $\varepsilon_{xx}$ and $\varepsilon_{yy}$ affect the exchange interactions, magnetic ground state, magnetocrystalline anisotropy, and magnetic ordering temperature in strongly anisotropic and composition-dependent ways.

The magnetic phase diagrams calculated using DFT show that the FM state remains stable over a significant region of strain space for all three monolayers, while sufficiently large and anisotropic deformations can stabilize competing Stripy, ZZ-1, and ZZ-2 antiferromagnetic configurations. The two Se-based compounds show the richest strain-induced phase competition, with all four magnetic configurations appearing within the investigated strain range. In contrast, MnSb$_2$Te$_4$ retains the FM ground state over a larger portion of strain space, indicating a more strain-insensitive ferromagnetic exchange hierarchy. Mapping the DFT energies onto a Heisenberg model further shows that this behavior arises from the collective and anisotropic change of $J_1$, $J_2$, and $J_3$ with strain.

The three compounds also show distinctly different responses of their magnetocrystalline anisotropy and finite-temperature magnetic ordering. MnSb$_2$Te$_4$ has a larger out-of-plane magnetic anisotropy than the two Se-based monolayers and, within the investigated strain range, this anisotropy remains comparatively insensitive to deformation. By contrast, the anisotropy of MnBi$_2$Se$_4$ and MnSb$_2$Se$_4$ is more readily modified by strain. The calculated magnetic ordering-temperature landscapes further demonstrate that the magnetic ordering temperature $T_{\mathrm{C}}$ does not simply follow the dominant nearest-neighbor interaction $J_1$. Instead, it reflects the collective balance among $J_1$, $J_2$, and $J_3$, together with the magnetic anisotropy that stabilizes long-range magnetic order in two dimensions. Consequently, strain can enhance $T_{\mathrm{C}}$ even when $J_1$ is weakened, provided that the accompanying redistribution of the longer-range interactions increases the overall stability of the ferromagnetic state. Among the three materials, MnSb$_2$Te$_4$ exhibits the broadest strain window over which elevated $T_{\mathrm{C}}$ and ferromagnetic ordering are simultaneously retained, making it particularly attractive for applications requiring tolerance to variations in the applied deformation.

We conclude by outlining several directions for future work. First, the configurations at the largest strains provide an opportunity to further explore structural, mechanical, and dynamical stability through elastic stability analysis and phonon calculations. Second, atom-, orbital-, and band-resolved analyses of the spin-orbit-coupling energy can provide microscopic insight into the distinct strain dependence of the magnetocrystalline anisotropy in the Se- and Te-based compounds. Third, since substrates provide a practical means of controlling strain in two-dimensional materials, DFT calculations that explicitly account for substrate effects can clarify how the intrinsic strain trends identified here translate to realistic device geometries. Ferroelectric substrates offer an additional degree of control over magnetic properties \cite{xue2020control}, potentially enabling mechanical strain to be combined with electrostatic or ferroelectric control of the magnetic state. It is also of interest to determine whether the predicted strain regimes can be accessed experimentally through epitaxial lattice mismatch, flexible substrates, piezoelectric coupling, or other approaches for controlled in-plane deformation. The effects of substrates, defects, carrier density, and finite sample thickness also provide promising directions for further investigation, as each can modify the exchange interactions and magnetic anisotropy relative to the isolated-monolayer limit considered here. More broadly, the strong coupling between lattice deformation and magnetic order identified in these calculations motivates further study of simultaneous strain control of magnetic and electronic properties in this family of materials. Finally, a material- and strain-dependent determination of $U$, analogous to recent DMC benchmarking for MnBi$_2$Te$_4$ \cite{ahn2026optimizing}, may further refine the quantitative phase boundaries and ordering temperatures predicted here. Such studies may provide a route toward mechanically tunable two-dimensional magnets in which the magnetic ground state, anisotropy, and ordering temperature are engineered through composition and directional strain.

\section*{Acknowledgement}
This research used resources of the Oak Ridge Leadership Computing Facility, which is a DOE Office of Science User Facility supported under Contract DE-AC05-00OR22725

During the preparation of this work the authors used OpenAI (GPT-5.6) in order to search literature, verify assumptions, correct grammar and revise language. After using this tool/service, the authors reviewed and edited the content as needed and takes full responsibility for the content of the published article.

\bibliographystyle{unsrt}
\bibliography{references}

\end{document}